\documentclass[aip,jcp,reprint,amsmath,amssymb,floatfix,citeautoscript,nofootinbib]{revtex4-1}
\usepackage{cancel}
\usepackage{amsmath}
\usepackage{physics}
\usepackage{graphicx}
\usepackage{amssymb}
\usepackage{amsthm}
\usepackage{bm}
\usepackage{dcolumn}% Align table columns on decimal point
\newcolumntype{x}[1]{D{.}{.}{#1}}

\usepackage{ragged2e}
\usepackage{txfonts}
\allowdisplaybreaks

\usepackage{color}
\definecolor{myblue}{rgb}{0,0,1}
\usepackage[breaklinks=true,colorlinks=true,linkcolor=myblue,urlcolor=myblue,citecolor=myblue]{hyperref}

\usepackage{mathtools}
\usepackage{braket}% Enables braket notation
\usepackage{longtable}
\usepackage[version=3]{mhchem}
\usepackage[capitalise]{cleveref}
\usepackage{siunitx}
\newcommand{\myred}[1]{{#1}}

\newcommand{\revision}[1]{{\myred{#1}}}
\newcommand{\revisiontcb}[1]{\myred{#1}}
\newcommand{\revisiontcbb}[1]{\myred{#1}}
\usepackage{ulem}
\newcommand{\alo}{Al\textsubscript{2}O\textsubscript{3}}
\newcommand{\alosurf}{$\alpha$-\alo{}(0001)}
\newcommand{\tio}{TiO\textsubscript{2}}
\newcommand{\tiosurf}{rutile TiO\textsubscript{2}(110)}

\begin{document}

\title{
\textit{Ab Initio} Surface Chemistry with Chemical Accuracy: Application to Water on Metal Oxides
}
%Chemistry of water on metal oxide surfaces with periodic gold-standard quantum chemistry
% The chemical fate of adsorbed water on metal oxide surfaces with periodic gold-standard quantum chemistry

\author{Hong-Zhou Ye}
\email{hzyechem@gmail.com}
\affiliation{Department of Chemistry, Columbia University, New York, NY 10027 USA}
\author{Timothy C. Berkelbach}
\email{t.berkelbach@columbia.edu}
\affiliation{Department of Chemistry, Columbia University, New York, NY 10027 USA}

\begin{abstract}
    First-principles calculations are a cornerstone of modern surface science and heterogeneous catalysis.
    However, accurate reaction energies and barrier heights are frequently inaccessible due to the approximations demanded by the large number of atoms.
    Here we \revisiontcbb{combine developments in local correlation and periodic correlated wavefunction theory} to solve the many-electron Schr\"odinger equation for molecules on surfaces with chemical accuracy, commonly defined as 1~kcal/mol.
    As a demonstration, we study water on the surface of \alo{} and \tio{}, two prototypical and industrially important metal oxides for which we obtain converged energies at the level of coupled-cluster theory with single, double, and perturbative triple excitations [CCSD(T)], commonly known as the ``gold-standard'' in molecular quantum chemistry.
    We definitively resolve the energetics associated with water adsorption and dissociation, enabling us to address recent experiments and to analyze the errors of more commonly used approximate theories.
\end{abstract}

\maketitle

\section{Introduction}

The structure, bonding, and chemistry of molecules and materials is governed by the
many-electron Schr\"odinger equation, which, for all but the simplest systems, must be solved approximately using
numerical techniques. Especially for solids and surfaces, containing a semi-infinite
number of atoms, severe approximations have historically been necessary,
and a primary effort of computational materials science has been the gradual
elimination of these approximations.
Early work used noninteracting and mean-field theories of band structure,
evolving into the popular density functional theory (DFT),
all of which reduce the many-electron Schr\"odinger equation to a set of self-consistent
one-electron Schr\"odinger equations.
However, the limitations of DFT have been noted in the context of chemical reactions~\cite{Cohen12CR,Mardirossian17MP},
surface adsorption~\cite{Schimka10NM,Araujo22NC}, and
heterogeneous catalysis~\cite{Libisch14ACR,Gaggioli19ACSCatal,Schafer21JPCL}, encouraging
the development of more accurate methods applicable to complex systems epitomized
by molecules on periodic solid surfaces.

To go beyond one-electron theories, explicit electron correlations can be reintroduced
with finite- or infinite-order perturbation theories that can, in principle, be systematically
converged to a numerically exact solution.
Here, we show that, with new methodological developments, this convergence can be achieved
along all necessary axes---including the description of electron correlation, the one-electron
basis set, and the size of the model surface---to provide surface chemistry energetics with
chemical accuracy, comparable to that which is achievable for small-molecule main-group
chemistry~\cite{Tajti04JCP,Curtiss07JCP}.  Specifically, as our highest level of theory,
we apply coupled-cluster theory with single, double, and perturbative triple
excitations [CCSD(T)]~\cite{Bartlett07RMP},
commonly known as the ``gold standard'' in molecular quantum chemistry.
The application of such methods to
solids can be traced back to 1990s~\cite{Stoll92PRB,Paulus06PR} and has been increasingly pursued over the last
few years.~\cite{Pisani12PCCP,Booth13Nature,Yang14Science,Zhang19FM,Lau21JPCL,Kubas16JPCL,Sauer19ACR,Brandenburg19JPCL,Maristella19JCTC,Schafer21JCPb,Mullan22JCP,Shi23JACS}
We leverage several recent developments in periodic integral evaluation~\cite{Ye21JCPa,Ye21JCPb,Bintrim22JCTC}
and Gaussian basis sets~\cite{Ye22JCTC} along with a new implementation of
periodic CCSD(T) with local correlation to enable a quantitative study of
\revisiontcb{single molecule adsorption and reaction chemistry}
on solid surfaces at this high level of theory.

\section{Computational Methods}

Within the Born-Oppenheimer approximation, the total electronic energy
is expressed as a sum of the Hartree-Fock energy and the correlation energy,
$E = E_0 + E_\mathrm{c}$.
\revisiontcbb{We first calculate the Hartree-Fock energy using large supercells with periodic boundary conditions.}
We then use a local, fragment-based approach wherein the correlation energy of a supercell
containing $N$ electrons is expressed
as a sum of contributions from all $N$ localized occupied orbitals $i$,
$E_\mathrm{c} = \sum_{i=1}^{N} E_\mathrm{c}^{(i)}$.
Importantly, each contribution $E_\mathrm{c}^{(i)}$ is evaluated independently
in a truncated set of occupied and unoccupied orbitals
that are optimized for local orbital $i$; specifically, we use local natural
orbitals (LNOs) from second-order perturbation theory~\cite{Rolik11JCP,Nagy17JCP} \revisiontcbb{with periodic boundary conditions}.
For the insulating materials studied
here, the number of LNOs needed for a target accuracy is independent
of the total system size.
Thus, the cost of calculating the total correlation energy grows only linearly with $N$
and each calculation of $E_\mathrm{c}^{(i)}$ is independent of all others,
which enables highly efficient simulation of large systems through parallel computing.
By increasing the number of LNOs, we converge to the exact \revisiontcbb{periodic} CCSD(T) energy at a fraction of the cost.
This low cost allows us to simulate periodic solids with supercells
containing over 100 atoms and almost 1000 electrons using high-quality correlation-consistent
one-electron basis sets~\cite{Ye22JCTC} and thus to reliably eliminate errors
stemming from incomplete basis sets, cluster models, or small simulation cells.
\revisiontcb{Details of our computational methods are provided in the Supplementary Material.}

\section{Water on metal oxides}

As a demonstrative application of these developments, we study the interaction between
water and solid metal oxides, which are two of the most abundant substances on Earth.
Understanding the chemistry of their interaction is important for myriad
technological applications, including electronics, catalysis, and corrosion~\cite{Bjorneholm16CR,Mu17CSR,Barry21CR}.
For example, semiconducting metal oxides such as TiO$_2$ are popular
photocatalysts for solar water splitting~\cite{Schneider14CR,Rousseau20NRM}.
The chemistry of water on metal oxide surfaces also serves as an important model
for general surface chemistry and heterogeneous catalysis, motivating
extensive experimental and theoretical research efforts~\cite{Hass98Science,Kirsch14JPCC,Mu17CSR,Diebold17JCP},
with experiments primarily performed using temperature-programmed desorption~\cite{Henderson96SS,Petrik18JPCC}, vibrational sum frequency generation~\cite{Tong15JCP},
and scanning probe microscopies~\cite{Brookes01PRL,Wang17PNAS}.
Specifically, we study \alo{} and \tio{}, two prototypical metal oxides and subjects of ongoing debates
about the fate of a molecularly adsorbed water
molecule~\cite{Petrik18JPCC,Wang17PNAS,Diebold17JCP},
which we aim to resolve in the present work.

We first consider \alo{}, which is a common support
in heterogeneous catalysis and has
been intensively studied as a model metal-oxide surface for water reactivity~\cite{Hass98Science,Ranea09JPCC,Kirsch14JPCC,Petrik18JPCC,Mullan22JCP}.
In particular, the most stable \alosurf{} surface has been
characterized, computationally by DFT and experimentally under
ultrahigh vacuum, to be aluminum-terminated with significant
structural distortions.

\begin{figure*}
    \centering
    \includegraphics[width=0.95\linewidth]{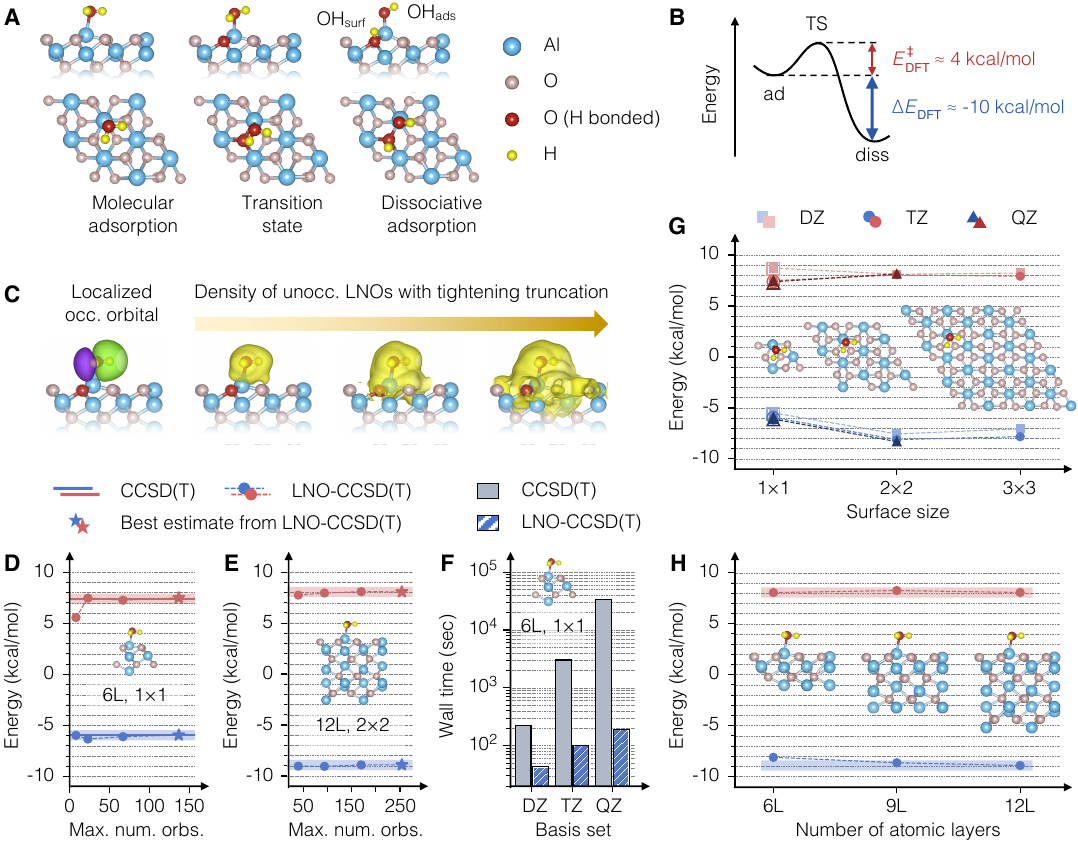}
    \caption{%
    (A) Atomic structure of a single water molecule adsorbed on the \alosurf{} surface.
    The molecularly adsorbed water molecule (left) may transfer a hydrogen to a neighboring surface oxygen via the transition state (middle), resulting in OH\textsubscript{ads} and OH\textsubscript{surf} fragments (right).
    (B) Schematic illustration of the potential energy surface associated with the water dissociation reaction predicted by DFT.
    % \tcb{Why use the DFT numbers here? What about leading the symbols but dropping the numbers? If you want to keep DFT, then maybe we should subscript the $E$ with ``DFT''.}
    (C) Isosurface visualization of a representative localized occupied orbital and the density of the corresponding unoccupied LNOs, the number of which increases with tightening truncation threshold (left to right).
    (D) The convergence of the reaction energy (blue) and barrier (red) calculated by LNO-CCSD(T) with the LNO subspace size for a small surface model of 6 atomic layers and $1 \times 1$ surface using a TZ basis set.
    With about 100 LNOs per occupied orbital, the LNO-CCSD(T) energies converge to the canonical CCSD(T) results (solid horizontal line) to an accuracy better than 1~kcal/mol (shaded area).
    (E) The same as in (D), but for a larger surface model with 12 atomic layers and $2 \times 2$ surface size, where canonical CCSD(T) is unavailable.
    (F) Wall time of LNO-CCSD(T) calculations for the 6L/$1\times 1$ surface model using basis sets of increasing size compared to canonical CCSD(T).
    (G) Reaction energy (blue) and barrier (red) calculated by LNO-CCSD(T) for a 6L slab with increasing surface size and basis set size.
    The canonical CCSD(T) results are also shown in hollow markers for the $1 \times 1$ surface.
    Both energies are converged to better than 1~kcal/mol of accuracy using a $2 \times 2$ surface and a TZ basis set.
    (H) Reaction energy (blue) and barrier (red) calculated by LNO-CCSD(T) in a TZ basis set for a $2 \times 2$ surface with increasing slab thickness.
    Both energies are converged to better than 1~kcal/mol of accuracy using a 12L model.}
    \label{fig:al2o3_conv}
\end{figure*}

Water undergoes molecular adsorption through an interaction between the water molecule's oxygen lone pair and a three-fold coordinated surface aluminum, after which it can potentially dissociate, transferring a hydrogen atom to a neighboring surface oxygen atom, yielding OH\textsubscript{ads} and OH\textsubscript{surf} fragments (Fig.~\ref{fig:al2o3_conv}A).
DFT calculations spanning 25 years~\cite{Hass98Science,Ranea09JPCC,Wang11JPCC,Wirth12JPCC} all predict that dissociation is favorable by about 10~kcal/mol, with a small barrier of about 4~kcal/mol, suggesting fast and complete dissociation within a few nanoseconds at room temperature (Fig.~\ref{fig:al2o3_conv}B).
The first experimental support for this process came in 2014 by observing signatures of surface hydroxyls in vibrational spectroscopy~\cite{Kirsch14JPCC}.
However, later experiments using vibrational spectroscopy and temperature-programmed desorption found that the dissociated products can take days to form even in ambient conditions~\cite{Tong15JCP,Petrik18JPCC}.
This timescale is in stark contrast to the small reaction barrier predicted by DFT, highlighting
the challenge of studying elementary chemical reactions on well-defined surfaces
and raising questions about the origin of this discrepancy.

With high-level periodic quantum chemistry methods in hand, we can accurately quantify the surface reaction energetics, which must be carefully converged with respect to the number of correlated orbitals, the basis set size, the surface size, and the slab thickness (Fig.~\ref{fig:al2o3_conv}C--H).
\revisiontcb{All geometries were optimized using DFT with the popular Perdew-Burke-Ernzerhof (PBE) functional~\cite{Perdew96PRL} and the D3 dispersion correction~\cite{Grimme10JCP} (PBE+D3), which was used by previous theoretical studies of the same system~\cite{Ranea09JPCC}, and further details are provided in the Supplementary Material.}
Figure~\ref{fig:al2o3_conv}C visualizes the unoccupied LNO subspace for a representative localized occupied orbital, where the number of LNOs can be increased systematically by tightening the truncation threshold.
Figure~\ref{fig:al2o3_conv}D shows the convergence of the LNO-CCSD(T) reaction energy and barrier height for a small $1\times 1$ surface model containing six atomic layers (6L/$1 \times 1$) with a triple-zeta (TZ) basis set, which is the largest system where canonical CCSD(T) results can be generated for comparison.
Both energies are seen to converge quickly to the canonical results within chemical accuracy by using less than 100 LNOs per occupied orbital, which is a small fraction of the total number of orbitals (about 630).
This fast convergence is consistent with the large gap of \alo{} and its weakly correlated, main-group electronic structure.
As shown in Fig.~\ref{fig:al2o3_conv}F, the smaller number of LNOs results in significant speedups of CCSD(T); for the large basis sets necessary to eliminate basis set incompleteness errors, the speedup is more than a factor of 100.
This high computational efficiency allows us to apply LNO-CCSD(T) to much larger surface models beyond the reach of canonical CCSD(T), as exemplified in Fig.~\ref{fig:al2o3_conv}E for a 12L/$2 \times 2$ surface model that contains over 80 atoms and 2000 orbitals in the TZ basis.
The LNO-CCSD(T) energies again show quick convergence, requiring a number of LNOs comparable to that of the smaller system, demonstrating the ability to scale to large systems without the significant increase in cost that accompanies canonical CCSD(T).
We are thus able to fully converge the reaction energy and barrier height with respect to the surface size, basis set size (Fig.~\ref{fig:al2o3_conv}G), and slab thickness (Fig.~\ref{fig:al2o3_conv}H).
\revisiontcb{We have used the same method to calculate the adsorption energy, using a counterpoise correction for basis set superposition error.}

\begin{figure*}
    \centering
    \includegraphics[width=0.8\linewidth]{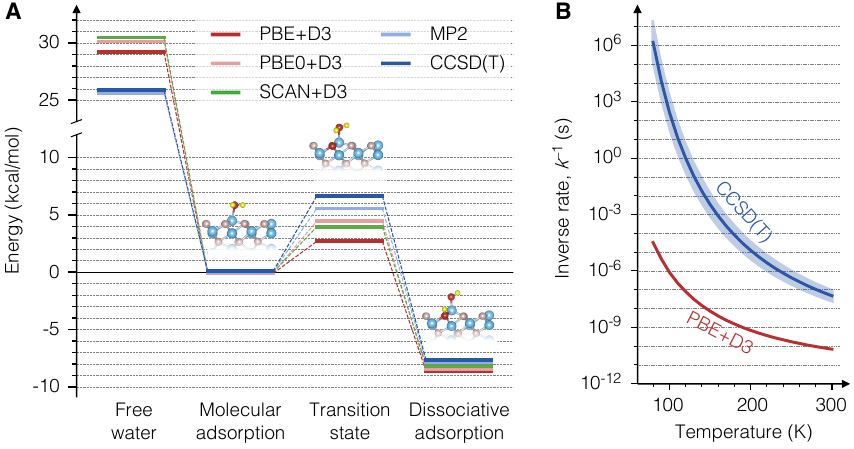}
    \caption{
    (A) \revision{Zero-point energy corrected} reaction energetics associated with the water \revision{adsorption and} dissociation reaction calculated using different electronic structure methods.
    (B) Inverse rate for the water dissociation reaction in the temperature range of 80~K to 300~K evaluated using harmonic transition state theory.
    For CCSD(T), the shaded area indicates an energy uncertainty of $\pm 0.5$~kcal/mol.
    }
    \label{fig:al2o3_main}
\end{figure*}

Our final results obtained using \revision{periodic LNO-}CCSD(T) are presented in Fig.~\ref{fig:al2o3_main}A, \revisiontcb{where they are compared to those obtained using several other levels of theory}.
\revision{All energies have been corrected by the vibrational zero-point energy (ZPE) evaluated using PBE}, \revisiontcb{which yields non-negligible corrections of $-0.4$~kcal/mol for the magnitude of the adsorption energy, $-0.6$~kcal/mol for the dissociation reaction energy, and $-2.2$~kcal/mol for the barrier height.}
\revision{CCSD(T) predicts a sizable adsorption energy of $26.5$~kcal/mol, while PBE+D3 overestimates this number by about $3$~kcal/mol.
This trend is consistent with that observed previously for water adsorption on the MgO (001) surface~\cite{Maristella19JCTC}.
}
CCSD(T) also confirms that the dissociative adsorption product predicted by PBE+D3 is thermodynamically more stable than the molecularly adsorbed water by about 8~kcal/mol.
\revisiontcb{The dissociation energy of molecularly adsorbed water is $-7.5$~kcal/mol from CCSD(T), which is in good agreement with the value of $-8.6$~kcal/mol from PBE+D3.}
However, the reaction barrier height from CCSD(T) is 6.6~kcal/mol, which is more than twice that predicted by PBE+D3 (2.8~kcal/mol, \revisiontcb{i.e., an underestimation by almost 4~kcal/mol}).

The underestimation of reaction barriers by DFT with a generalized gradient approximation (GGA) functional like PBE is well-known and can be traced back to the systematic self-interaction error (SIE) of semilocal functionals~\cite{Cohen12CR}.
In Fig.~\ref{fig:al2o3_main}A, we also show the reaction energetics obtained using DFT with a \revision{a meta-GGA functional, SCAN~\cite{Sun15PRL}+D3}, a hybrid functional, PBE0~\cite{Adamo99JCP}+D3, and the second-order M{\o}ller-Plesset perturbation theory, MP2.
\revisiontcb{The two functionals mitigate the SIE by satisfying more exact constraints on the exchange-correlation functional~\cite{Sun15PRL} (SCAN) or through partial inclusion of the exact exchange energy~\cite{Mardirossian17MP} (PBE0)}, while MP2 completed removes the SIE and includes approximate many-body electron correlation to second order.
All three methods are seen to improve upon PBE+D3 in the calculated reaction barrier, \revisiontcb{predicting barriers of 4.0~kcal/mol (SCAN+D3), 4.5~kcal/mol (PBE0+D3), and 5.6~kcal/mol (MP2)}.
\revisiontcb{However, SCAN+D3 and PBE0+D3 show even stronger overestimation of the adsorption energy compared to PBE+D3, highlighting the challenge of designing a universal functional for surface chemistry, while the wavefunction-based MP2 theory agrees quantitatively with CCSD(T).}

The higher barrier predicted by CCSD(T) has a significant impact on the kinetics of the surface reaction over a wide range of temperature.
We approximate the reaction rate using harmonic transition state theory~\cite{Peter17Book},
$k(T) = (k_{\textrm{B}}T/h) \exp\left\{-\left[\Delta E^{\ddagger} + \Delta F^{\ddagger}_{\textrm{vib}}(T)\right]/k_{\textrm{B}}T\right\}$,
where $h$ is Planck's constant, $k_{\textrm{B}}$ is Boltzmann's constant, $T$ is temperature, $\Delta E^{\ddagger}$ is the reaction barrier \revision{without ZPE correction},
and $\Delta F_{\textrm{vib}}^{\ddagger}(T)$ is the temperature-dependent vibrational activation free energy, which we \revision{calculated using harmonic frequencies from PBE} \revisiontcb{(the vibrational correction to the barrier height is reduced in magnitude from $-2.2$~kcal/mol at 0~K to $-1.4$~kcal/mol at 300~K; see Fig.~S4 in the Supplementary Material)}.
The reaction timescales using CCSD(T) and PBE+D3 barrier heights are shown in Fig.~\ref{fig:al2o3_main}B for the temperature range of 80--300~K.

At 300~K, all levels of theory predict the dissociation of the water O-H bond to be fast on the surface, with CCSD(T) predicting $k^{-1} \approx$~\SI{0.1}{\micro\second}, which is three orders of magnitude slower than $k^{-1} \approx 0.1$~ns from PBE+D3.
The difference between CCSD(T) and PBE+D3 becomes even more prominent at lower temperature.
% At 100~K, PBE+D3 still predicts fast dissociation with $k^{-1} \approx 1$~ms.
At \revision{80--100~K}, PBE+D3 still predicts fast dissociation with $k^{-1} \approx$~\revision{\SI{1}{\micro\second}}.
By contrast, CCSD(T) predicts slow kinetics with a time scale of about a day, i.e., about nine orders of magnitude slower than that predicted by PBE+D3.
While the slow kinetics at about 100~K predicted by the CCSD(T) barrier height are consistent with
cryogenic data~\cite{Petrik18JPCC}, the prediction at 300~K does not agree with experiments performed at room temperature~\cite{Tong15JCP}.

Although classical transition state theory is approximate, due to its neglect of recrossing events and quantum tunnelling~\cite{Peter17Book}, it is highly unlikely that these corrections would be large enough~\cite{Wirth12JPCC} to predict a lifetime consistent with room-temperature experiments~\cite{Tong15JCP}, especially in ultrahigh vacuum.
Therefore, we conclude that unimolecular adsorption and dissociation of water
on a pristine \alosurf{} surface requires about a day at 100~K, but should occur by this mechanism
on the sub-microsecond time scale at 300~K.
We also predict the dissociation to be essentially irreversible: at 300~K,
we predict $k^{-1} \approx 10$~s for the recombination of the two surface hydroxyl groups, which is $10^8$ times slower than the dissociation reaction.
Deviations seen in experiment must be associated with mechanisms not considered
here, such as alternative surface motifs, competing reaction pathways, or
cooperative effects~\cite{Petrik18JPCC}.
\revisiontcb{However, even in the presence of alternative pathways, our predicted reaction timescale should be an upper bound to the
one observed, and so understanding why our studied reaction path is not the one realized in experiments is unclear.}
\revisiontcb{For our studied reaction path,} inaccuracies of the electronic structure theory, which we have shown to be large with common
density functional approximations, have been eliminated by our work.
\revisiontcb{Within the approximations of transition state theory, the reaction barrier needed to explain a reaction timescale on the order of days at room temperature is more than 25~kcal/mol; for CCSD(T) to be so wrong would be unprecedented.}

\begin{figure*}[!t]
    \centering
    \includegraphics[width=0.95\linewidth]{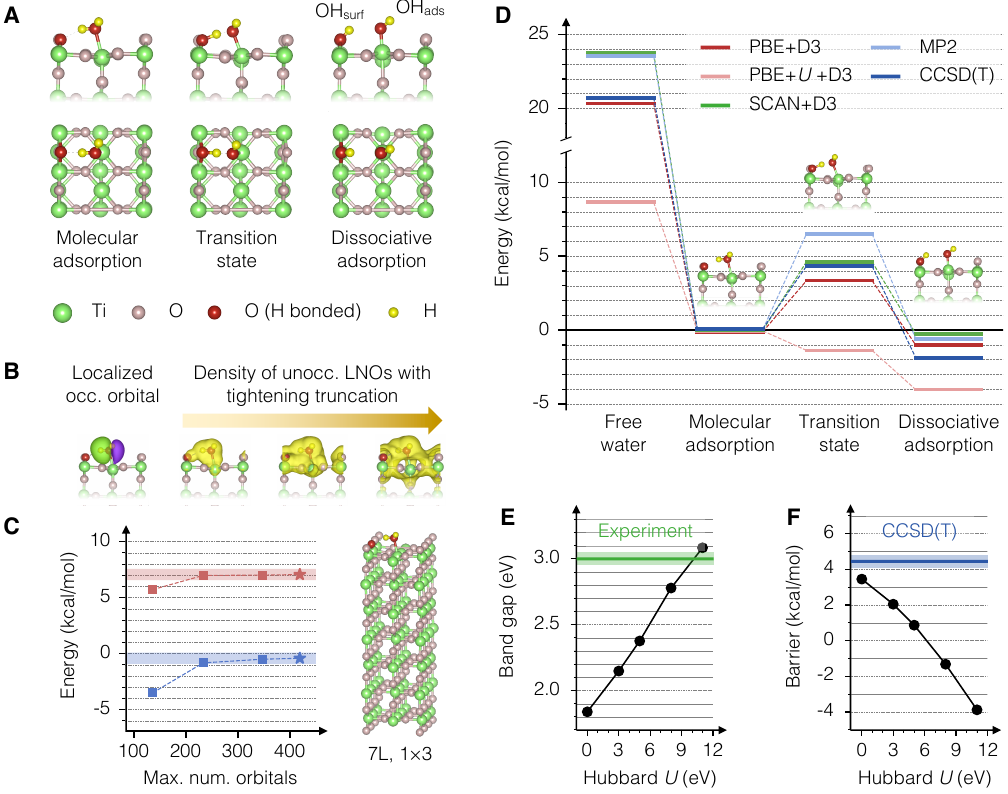}
    \caption{
    (A) Atomic structure of a single water molecule adsorbed on the \tiosurf{} surface (left), which may transfer a hydrogen to a neighboring surface oxygen via the transition state (middle), leaving OH\textsubscript{ads} and OH\textsubscript{surf} fragments (right).
    (B) Isosurface visualization of a representative localized occupied orbital and the density of the corresponding unoccupied LNOs obtained with tightening truncation (left to right).
    (C) The convergence of the LNO-CCSD(T) reaction energy (blue) and barrier (red) with the LNO subspace size for a 7L/$1 \times 3$ model using a TZ basis set.
    Both energies are converged to an accuracy better than 1~kcal/mol (shaded area) with about 300 LNOs per occupied orbital, which is a small fraction of the total orbital count (about 5000).
    (D) \revision{ZPE-corrected} reaction energetics associated with the water \revision{adsorption and} dissociation reaction calculated using different electronic structure methods.
    For PBE$+U$, $U = 8$~eV is applied to the $3d$ band of Ti as suggested by previous work~\cite{Dompablo11JCP}.
    % \revision{The available experimental heat of adsorption~\cite{Hugenschmidt94SS,Campbell13CR} and reaction energetics~\cite{Wang17PNAS} are included for comparison.}
    (E) Band gap of bulk rutile \tio{} predicted by PBE$+U$ with different values of $U$, compared with the experimental value~\cite{Cronemeyer52PR}.
    (F) \revision{ZPE-corrected} barrier height for the water dissociation predicted by PBE$+U$+D3 with different values of $U$, compared with the CCSD(T) reference.
    }
    \label{fig:tio2_main}
\end{figure*}

\revisiontcb{To further explore the performance of CCSD(T) for surface reactions, we study} \tio{}, which has a more
complicated electronic structure due to the 3d electrons of the transition metal Ti.
The water-\tio{} interface has been a focus of intensive research activities since the 1970s for its importance in photocatalytic water splitting for hydrogen generation~\cite{Schneider14CR,Rousseau20NRM}.
Particularly, the most stable \tiosurf{} surface has been characterized, computationally by DFT and experimentally under ultrahigh vacuum, to be terminated by alternating rows of five-fold coordinated titaniums (Ti\textsubscript{5c}) and bridging oxygens (O\textsubscript{b}).
Water undergoes molecular adsorption via interaction of its lone pair with
Ti\textsubscript{5c} and a weak hydrogen bond with a neighboring
O\textsubscript{b}.

Like for \alo{}, the possibility of subsequent dissociative adsorption is debated---not
only the kinetics but also the thermodynamics.
The hypothesized dissociation occurs through water transferring a proton to its neighboring O\textsubscript{b}, yielding OH\textsubscript{ads} and OH\textsubscript{surf} fragments (Fig.~\ref{fig:tio2_main}A).
Controversial results were seen in early experimental studies, wherein temperature-programmed desorption~\cite{Hugenschmidt94SS} and scanning tunneling microscope~\cite{Brookes01PRL,Schaub01PRL} experiments found signals of only molecularly adsorbed water in the absence of surface defects, while photoemission spectroscopy~\cite{Duncan12PRB} observed signatures of OH\textsubscript{surf} and supported mixed molecular and dissociative adsorption.
In a recent combined scanning tunneling microscope-molecular beam experiment, dissociative adsorption on the Ti\textsubscript{5c} site was measured at very low water coverage to be slightly higher in energy than molecular adsorption by 0.8~kcal/mol, with a barrier height of 8.3~kcal/mol~\cite{Wang17PNAS}.
Early DFT calculations found conflicting predictions on the relative stability of the two adsorption states~\cite{Schaub01PRL,Harris04PRL,Lindan05PRB},
where results depend sensitively on choices of approximate functionals~\cite{Lindan05PRL,Harris05PRL}, dispersion corrections~\cite{Kumar13JPCC}, and treatments of strong electron correlation~\cite{Dompablo11JCP,Kumar13JPCC}.
\revision{More recent DFT-based studies employing the SCAN functional along with machine learning potentials to enable molecular dynamics simulation found that dissociative adsorption is higher in energy by 0.5--1~kcal/mol~\cite{Wen23PNAS,Zeng23NC}, in agreement with the recent experiment~\cite{Wang17PNAS}.
However, the barrier height was predicted to be about $6$~kcal/mol~\cite{Zeng23NC}, which is more than 2~kcal/mol lower than the experiment.}

We revisit this problem with \revision{periodic LNO-}CCSD(T) for an isolated water molecule on the \tiosurf{} surface.
Despite the more strongly correlated electronic structure of a transition metal oxide, the LNOs are well-localized around their associated localized occupied orbital (Fig.~\ref{fig:tio2_main}B).
We have performed a convergence study analogous to that for \alo{} and determined that, \revisiontcb{based on PBE+D3 calculations (Fig.~S6)}, the surface chemistry energetics can be converged using a 7L/$1\times 3$ surface model (Fig.~\ref{fig:tio2_main}C) that contains about 130 atoms and 5000 orbitals when using a TZ basis set and that the CCSD(T) calculations require about 300 LNOs per occupied orbital.
Our final \revision{ZPE-corrected adsorption and reaction energetics} are presented in Fig.~\ref{fig:tio2_main}D, where they are compared to DFT using PBE+D3 and SCAN+D3.

Our CCSD(T) calculations predict a ZPE-corrected adsorption energy of $20.8$ kcal/mol, which is in excellent agreement with the 20--22~kcal/mol heat of adsorption determined by temperature-programmed desorption experiments~\cite{Hugenschmidt94SS,Campbell13CR}.
Our adsorption energy is also in good agreement with a previous local CCSD(T) calculation that used an embedded cluster approach~\cite{Kubas16JPCL}, but it is smaller in magnitude, by about 8~kcal/mol, than a more recent periodic CCSD(T) calculation that used a local embedding approach~\cite{Schafer21JCPa}.
\revision{Both SCAN+D3 and MP2 overestimate our CCSD(T) energy by $3$~kcal/mol, while PBE+D3 agrees with CCSD(T) within chemical accuracy.
}

\revision{For the surface water dissociation, CCSD(T) predicts dissociative adsorption to be slightly lower in energy than molecular adsorption by 1.9~kcal/mol, with a barrier height of $4.3$~kcal/mol.
Compared to the recent experimental estimation~\cite{Wang17PNAS}, these numbers correspond to an overstabilization of the dissociation product by about 3~kcal/mol and an underestimation of the barrier height by about 4~kcal/mol.
}
\revisiontcb{Like for the case of water on \alo{}, the comparison between calculations and experiment is clearly a challenge, which we will discuss further in the Conclusions, but by taking the CCSD(T) results as a reference, we can evaluate the accuracy of lower levels of theory.}
\revisiontcb{Both PBE+D3 and SCAN+D3 yield a reaction profile similar to that of CCSD(T), predicting that dissociation is favorable by $1$~kcal/mol from PBE+D3 and $0.2$~kcal/mol from SCAN+D3, and a barrier height of $3.4$~kcal/mol from PBE+D3 and $4.6$~kcal/mol from SCAN+D3.}
By contrast, MP2 overestimates the barrier by about $2$~kcal/mol and underestimates the dissociation energy by about $1.5$~kcal/mol, suggesting that third- and higher-order electron correlation effects play an important role in the water-\tio{} chemistry.

A common approach to improving the performance of semilocal functionals like PBE,
especially in transition-metal containing materials, is the DFT+$U$
method~\cite{Dudarev98PRB}. For \tio{}, this method introduces
a local Coulomb interaction to the Ti $3d$ atomic orbital, raising the energy of the
Ti $3d$ band and yielding an increased band gap in better agreement with experiment.
For bulk rutile \tio{}, the experimental band gap is about 3.0~eV and PBE predicts
1.8~eV; PBE+$U$ with the value $U=8$~eV predicts 2.9~eV~\cite{Dompablo11JCP}.
Applying the same approach to the water dissociation reaction, we find that it
worsens agreement with CCSD(T): it predicts an exoergic reaction
by about $4.0$~kcal/mol and a negative reaction barrier of $-1.4$~kcal/mol (Fig.~\ref{fig:tio2_main}D).
This behavior can be understood by considering the bonding of the molecularly
adsorbed water molecule. Raising the energy of the empty accepting Ti $3d$ orbital
weakens the bond and destabilizes the molecularly adsorbed water, hence leading to a smaller barrier height and a more exoergic reaction.
\revision{This also explains the significantly smaller adsorption energy predicted by PBE+$U$+D3 ($8.7$~kcal/mol), also shown in Fig.~\ref{fig:tio2_main}D.}

In Fig.~\ref{fig:tio2_main}E and F, we show the bulk band gap and the water dissociation barrier
height as a function of the parameter $U$.
Improving one worsens the other, and there is no single value of $U$ that accurately predicts both properties.
This competing behavior is typical of DFT~\cite{Schimka10NM} and highlights the challenge of using DFT for computational heterogeneous catalysis, where stretched bonds, dispersion, and strong correlation are all important.
By contrast, high-level quantum chemical wavefunction approaches, such as CCSD(T), offer balanced and unbiased treatments of many-body electron correlation without empirical parameters.

\section{Conclusions}

To summarize, we computationally investigated the chemistry of water on two
prototypical metal-oxide surfaces, \alo{} and \tio{}, using state-of-the-art
periodic quantum chemistry at the CCSD(T) level of theory.
The quantitative reaction energetics made possible by this work sheds new light on
the long-standing puzzles between previous computational studies and
experimental observations regarding the chemical equilibrium and kinetics of
water molecules on these surfaces.
\revisiontcb{Perhaps most significantly, we have shown that accurate barrier heights and reaction energies based on ZPE-corrected single-point energies do not provide quantitative agreement with those inferred from experiments.
With CCSD(T), we believe that inaccuracies in the electronic structure are no longer responsible for the disagreement.
Instead, we propose that reconciling calculations and experiments requires a consideration of more complex reaction networks, thermodynamic and dynamic effects not captured by the harmonic approximation and transition state theory, or a reevaluation of experiments.
}
The results of our high-level quantum chemistry calculations also allowed
an unbiased examination of the performance of DFT for elementary steps of reactions
on real surfaces, for which comparisons to experiment are extremely challenging.
Such accurate calculations can immediately be used for validation, selection, or design
of more affordable density functional approximations or as training data for
machine learning of force fields~\cite{Noe20ARPC,Chen23JCTC,Yu23JPCL},
which could be used in molecular dynamics simulations to explore thermodynamic effects not captured by single-point calculations.~\cite{Wen23PNAS,Zeng23NC}
Although alternatives to CCSD(T) for multiconfigurational
electronic structure of strongly correlated solids must be pursued~\cite{Gaggioli19ACSCatal},
we anticipate that high-level periodic quantum chemistry approaches
will play an increasingly important role in the toolbox of surface science.
The large system sizes needed for convergence essentially demand the use of local correlation methods, such as the one used here.

\section*{Acknowledgements}
We thank Garnet Chan, Sandeep Sharma, and Yifan Li for useful discussions.
T.C.B.~acknowledges the hospitality of the Center for Computational Quantum Physics,
Flatiron Institute, where a portion of this work was completed.
The Flatiron Institute is a division of the Simons Foundation.
This work was supported by the National Science Foundation under
Grant Nos.~OAC-1931321 and CHE-1848369.
We acknowledge computing resources from Columbia University's Shared Research
Computing Facility project, which is supported by NIH Research Facility
Improvement Grant 1G20RR030893-01, and associated funds from the New York State
Empire State Development, Division of Science Technology and Innovation
(NYSTAR) Contract C090171, both awarded April 15, 2010.
Crystal structures and electronic orbitals were visualized using VESTA.~\cite{Momma11JAC}

\bibliography{refs}

\end{document}

\maketitle

    \vspace{3em}

    \hspace{2em}Note: figures and equations appearing in the main text will be referred to as
``Fig.~Mxxx'' and ``Eq.~Mxxx'' in this Supplementary Material document.

    \clearpage

    %\doublespacing
    \tableofcontents
    %\singlespacing

    \clearpage

    \section{Overall computational details}

    \subsection{Geometries}

    All geometries are optimized at the PBE(+D3) level of theory using the Quantum Espresso (QE) v6.4 code~\cite{Giannozzi09JPCM,Giannozzi17JPCM} with an energy convergence threshold of $0.2$~meV/cell and a force convergence threshold of $0.05$~eV/\AA{}.
    The projector augmented wave pseudopotentials for PBE as obtained from the QE official site are used for all elements (\cref{tab:QEPP}).
    For both surface systems (\alo{} and \tio{}), bulk geometries are first optimized, from which clean surfaces are constructed and optimized.
    The molecularly and dissociatively adsorbed water-surface geometries are then constructed and optimized, followed by climbing image nudged elastic band (CI-NEB) calculations to find the transition-state (TS) geometries.
    The D3 dispersion correction by Grimme and co-workers~\cite{Grimme10JCP} is used for the PBE geometry optimization of both clean surfaces and surfaces with adsorbates but not for bulk, \revision{where we found the effect of the dispersion correction to be small (see \cref{subsec:bulk_lat_const}).
    Other parameters affecting the geometry relaxation, including the plane-wave kinetic energy cutoff, the $k$-point mesh for Brillouin zone sampling, and the size of the slab model are detailed in \cref{sec:alo} for the \alo{} surface and \cref{sec:tio} for the \tio{} surface, respectively.
    The final optimized geometries are provided in \cref{subsec:opt_geom}.
    We also estimate the effect of the employed PBE+D3 geometries on the calculated adsorption and reaction energetics in \cref{subsec:est_geom_uncertainty}.
    }

    \begin{table}[!htbp]
        \centering
        \caption{The pseudopotential file (as obtained from the Quantum Espresso official site) and valence electrons for each element used by the plane-wave DFT calculations in this work.}
        \label{tab:QEPP}
        \begin{tabular}{lll}
            \hline
            \hline
            Element & PP file & Valence electrons  \\
            \hline
            Al & \verb|Al.pbe-n-kjpaw_psl.1.0.0.UPF|    & $3s^2 3p^1$ \\
            Ti & \verb|Ti.pbe-spn-kjpaw_psl.1.0.0.UPF|  & $3s^2 3p^6 3d^2$ \\
            O  & \verb|O.pbe-n-kjpaw_psl.1.0.0.UPF|     & $2s^2 2p^4$ \\
            H  & \verb|H.pbe-kjpaw_psl.1.0.0.UPF|       & $1s^1$ \\
            \hline
        \end{tabular}
    \end{table}

    \subsection{SCAN+D3, PBE+$U$+D3, PBE0+D3, HF, and MP2}

    Single-point calculations are performed at the PBE+D3-optimized geometries from above using different levels of theory to obtain the adsorption energy, reaction energies, and barrier heights reported in the main text.
    \revisiontcb{These single-point calculations were performed using a combination of QE and the PySCF package~\cite{Sun15JCC,Sun18WIRCMS,Sun20JCP}.}
    All numbers are carefully converged to within $1$~kcal/mol with respect to the basis set size, the surface size, the slab size, and Brillouin zone sampling, as detailed in \cref{sec:alo,sec:tio} for the two surface systems, respectively.
    The D3 dispersion correction is included in all reported DFT energies (PBE, SCAN, PBE0, and PBE+$U$).
    \revision{For PBE and PBE+$U$, we use the D3 energy for PBE calculated from QE.
    For SCAN (whose D3 parameters are not available in QE) and PBE0 (for which we performed calculations with PySCF), we calculate the respective D3 energies using the \texttt{simple-dftd3} code (\href{https://github.com/dftd3/simple-dftd3}{https://github.com/dftd3/simple-dftd3}), and we verified that this code gives the same D3 energy for PBE compared to QE.
    }

    The PBE$+U$ calculations are performed using QE with the simplified version of Cococcioni and de Gironcoli~\cite{Cococcioni05PRB} (\verb|lda_plus_u_kind = 0|) and the atomic projector (\verb|U_projection_type = 'atomic'|).
    The SCAN calculations are performed using QE with the norm-conserving Hartwigsen-Goedecker-Hutter (HGH) pseudopotentials~\cite{Goedecker96PRB,Hartwigsen98PRB} optimized for PBE.
    All other calculations, including the wavefunction calculations and the hybrid DFT calculations with the PBE0 functional, are performed using the PySCF code~\cite{Sun15JCC,Sun18WIRCMS,Sun20JCP} with translational symmetry-adapted Gaussian basis functions.
    The Goedecker-Teter-Hutter (GTH) pseudopotentials optimized for HF~\cite{HutterPP} or PBE~\cite{Goedecker96PRB,Hartwigsen98PRB} (for PBE0 only) are employed, with the same number of valence electrons as listed in \cref{tab:QEPP}.
    The correlation-consistent Gaussian basis set series optimized for the GTH pseudopotentials from Ref.~\citenum{Ye22JCTC}, GTH-cc-pV$X$Z with $X =$~D, T, Q (henceforth denoted by $X$Z), are used for all elements except for Ti, for which we performed similar basis optimization as in Ref.~\citenum{Ye22JCTC} to generate the corresponding GTH-cc-pV$X$Z series.
    We verified at PBE level that the reaction energetics calculated from PySCF agree within $1$~kcal/mol with those from QE when using a TZ basis set.
    The four-center electron-repulsion integrals are handled using the fast periodic density fitting algorithm from Ref.~\citenum{Ye21JCPa,YE21JCPb,Bintrim22JCTC}.
    The fitting basis sets are optimized for the GTH-cc-pV$X$Z series of these elements by minimizing the error of HF and MP2 energies of molecular monohydrides.
    We verified that the error in the calculated reaction energetics (from both mean-field and correlated calculations) introduced by these optimized fitting basis sets are less than $0.1$~kcal/mol by comparing to a large even-tempered fitting basis set generated by PySCF.
    All basis set data can be found in \url{https://github.com/hongzhouye/supporting_data/tree/main/2023/arXiv%3A2309.14640}.

    \revisiontcb{For HF and PBE0 calculations}, the finite-size error of the exact, nonlocal exchange energy arising from the integrable divergence at $\bm{G} = \bm{0}$ is treated using a Madelung constant correction~\citenum{Paier06JCP,Broqvist09PRB,Sundararaman13PRB}.
    This yields an energy with an asymptotic $1/N_k$ convergence to the thermodynamic limit (TDL), with $N_k$ the number of $k$-points used to sample the Brillouin zone.
    These ``finite-size error-corrected'' orbital energies are used to obtain both the MP2 correlation energy and the (T) part of the CCSD(T) correlation energy.
    \revisiontcb{In contrast, CCSD calculations are independent of the use of Madelung constant corrections, as long as they are consistently applied.}

    \subsection{LNO-CCSD(T)}
    \label{subsec:lnocc_theory}

    LNO-CCSD(T) calculations are performed using an in-house implementation in a developer version of PySCF based on the molecular LNO-CCSD(T) theory by K\'{a}llay and co-workers~\cite{Rolik11JCP,Rolik13JCP,Nagy17JCP}.
    In PySCF, most methods implemented for molecular calculations are automatically applicable to solids with $\Gamma$-point Brillouin zone sampling by replacing the preceding molecular mean-field calculations (HF in our case) with the corresponding periodic version.
    For the general case where the Brillouin zone is sampled by an evenly spaced, $\Gamma$-included $k$-point mesh, we first transform the $k$-point HF orbitals in the original simulation cell into the corresponding $\Gamma$-point HF orbitals in the BvK supercell and then perform LNO-CCSD(T) calculations therein.

    For completeness, we briefly review the LNO-CCSD(T) theory here and guide the reader to the original publications~\cite{Rolik11JCP,Rolik13JCP,Nagy17JCP} for details.
    All systems studied in this work are closed-shell and described by a spin-restricted HF reference determinant $\ket{\Phi_0}$, with canonical HF orbitals $\psi_p$, orbital energies $\epsilon_p$, and total energy $E_\mathrm{HF}$.
    We use $i,j,k$ for $N_\mathrm{o}$ occupied orbitals, $a,b,c$ for $N_\mathrm{v}$ virtual orbitals, and $p,q,r,s$ for $N$ unspecified molecular orbitals.
    In this basis, the electronic Hamiltonian is
    \begin{equation}
        H
            = \sum_{pq,\sigma}^{N} h_{pq} a_{p\sigma}^{\dagger} a_{q\sigma} +
            \frac{1}{2} \sum_{pqrs,\sigma\sigma'}^{N} V_{prqs} a_{p\sigma}^{\dagger} a_{q\sigma'}^{\dagger} a_{s\sigma'} a_{r\sigma}
            %= H_1 + H_2
    \end{equation}
    with $V_{pqrs} = (pq|rs)$ in $(11|22)$ notation.

    We choose the generalized Pipek-Mezey method~\cite{Lehtola14JCTC} with a \verb|minao| projector to localize the HF occupied orbitals
    \begin{equation}
        \phi_I
            = \sum_{i}^{N_{\mathrm{o}}} U_{iI} \psi_i
    \end{equation}
    Representative localized occupied orbitals are shown in Fig.~\fakeref{M1C} for water-\alo{} and \fakeref{M3B} for water-\tio{}.
    For each localized occupied orbital, $\phi_I$, one constructs a local active space $\mathcal{P}_I$ by augmenting $\phi_I$ with selected LNOs from second-order M{\o}ller-Plesset perturbation theory~\cite{Moller34PR} (MP2).
    Specifically, one computes the occupied-occupied and the virtual-virtual blocks of the MP2
    density matrix
    \begin{align}
        D^{I}_{ij}
            &= \sum_{ab} t^{(1)}_{iaIb} \big[ 2 t^{(1)}_{jaIb} - t^{(1)}_{Iajb} \big]
            \label{eq:DIij}  \\
        D^{I}_{ab}
            &= \sum_{jc} t^{(1)}_{Iajc} \big[ 2 t^{(1)}_{Ibjc} - t^{(1)}_{jbIc} \big]
            \label{eq:DIab}
    \end{align}
    where
    \begin{equation}    \label{eq:lno_approx_MP2t}
        t^{(1)}_{Iajb}
            = \frac{(Ia|jb)}{\tilde{\epsilon}_I + \epsilon_{j} - \epsilon_{a} - \epsilon_{b}}
    \end{equation}
    is an approximate MP2 amplitude with $\tilde{\epsilon}_I = \braket{\phi_I| f |\phi_I}$,
    and $f$ is the Fock operator.
    Diagonalizing the virtual-virtual block
    \begin{equation}
        D^{I}_{ab}
            = \sum_{c} \xi_c^{I} X_{ac}^{I} X_{bc}^{I},
    \end{equation}
    gives the virtual LNOs associated with $\phi_I$, i.e., $\phi_b = \sum_c X_{cb}^I\psi_c$.
    For the occupied LNOs, we follow Ref.~\citenum{Rolik11JCP} and diagonalize
    \begin{equation}    \label{eq:lno_occ}
        \tilde{D}^{I}_{ij}
            = \sum_{kl} Q_{ik}^{I} D_{kl}^{I} Q_{lj}^{I}
            = \sum_{k}^{N_{\textrm{o}}-1} \xi_k^{I} X_{ik}^{I} X_{jk}^{I}
    \end{equation}
    where $Q_{ij}^{I} = \delta_{ij} - U_{iI} U_{jI}$ projects out $\phi_I$ from $D_{ij}^{I}$ to prevent it from mixing with other occupied orbitals, giving the occupied LNOs $\phi_j = \sum_k X_{kj}^I \psi_k$.
    The eigenvalues $\xi^I_p$, which are between 2 and 0, quantify the importance of a given LNO to the electron correlation of localized orbital $\phi_I$.
    In practice, we construct the local active space $\mathcal{P}_I$ by keeping those LNOs satisfying
    \begin{equation}    \label{eq:lno_truncation}
        \xi_i \geq \lambda_{\textrm{o}},
        \quad{}
        \xi_a \geq \lambda_{\textrm{v}},
    \end{equation}
    for some user-selected thresholds $\lambda_{\textrm{o}}$ and $\lambda_{\textrm{v}}$, producing $n=n_\mathrm{o}+n_\mathrm{v}$ orbitals.
    Typically, $n$ is much less than $N$ and, for gapped systems such as the surface systems studied in this work, does not increase with system size for a targeted level of accuracy.
    In Fig.~\fakeref{M1C} and \fakeref{M3B}, we also show the density of the virtual LNOs generated for the representative localized occupied orbitals with $\lambda_v = 10^{-5}$, $10^{-6}$, and $10^{-7}$.

    A local Hamiltonian is then constructed by projecting $H$ into $\mathcal{P}_I$
    \begin{equation}    \label{eq:lno_HI}
        H_{I}
            = \sum_{pq\in \mathcal{P}_I,\sigma}^n
            f^{I}_{pq} a_{p\sigma}^{\dagger} a_{q\sigma}
            + \frac{1}{2} \sum_{pqrs \in \mathcal{P}_I,\sigma\sigma'}^n V_{prqs}
            a_{p\sigma}^{\dagger} a_{q\sigma'}^{\dagger} a_{s\sigma'} a_{r\sigma'}
    \end{equation}
    where
    \begin{equation}
        f_{pq}^{I}
            = h_{pq} + \sum_{j \notin \mathcal{P}_I}^{N_\mathrm{o}-n_\mathrm{o}} \left(2V_{pqjj} - V_{pjj q}\right),
    \end{equation}
    which includes a frozen core contribution.
    Solving the CCSD amplitude equations with $H_I$ gives the local CCSD amplitudes $t_{ia}$ and $t_{iajb}$ in $\mathcal{P}_I$, from which the local contribution of the CCSD(T) correlation energy $E^{(I)}$ can be evaluated by
    \begin{equation}
        E^{(I)}_{\mathrm{CCSD(T)}}
            = \sum_{kl \in \mathcal{P}_I}
            U_{kI} \left[ G^{(I),\mathrm{CCSD}}_{kl} + G^{(I),\mathrm{(T)}}_{kl} \right] U_{lI}
    \end{equation}
    where
    \begin{equation}
        G^{(I),\mathrm{CCSD}}_{kl}
            = \sum_{iab \in \mathcal{P}_I} (t_{iakb} + t_{ia} t_{kb}) (2 V_{ialb} - V_{ibla})
    \end{equation}
    and
    \begin{equation}
    \begin{split}
        G^{(I),\mathrm{(T)}}_{kl}
            &= -\frac{1}{3} \sum_{a \geq b \geq c \in \mathcal{P}_I} f_{abc}
            \sum_{ij \in \mathcal{P}_I} \bigg\{ \\
            &w_{ijk}^{abc}   \left( +8 v_{ijl}^{abc} -5 v_{ikj'}^{abc} -2 v_{jil}^{abc} +2 v_{jli}^{abc} +2 v_{lij}^{abc} -5 v_{lji}^{abc} \right) + \\
            &w_{ikj}^{abc}   \left( -5 v_{ijl}^{abc} +8 v_{ilj}^{abc} +2 v_{jil}^{abc} -2 v_{jli}^{abc} -5 v_{lij}^{abc} +2 v_{lji}^{abc} \right) +    \\
            &w_{kij}^{abc}   \left( +2 v_{ijl}^{abc} -5 v_{ilj}^{abc} -5 v_{jil}^{abc} +2 v_{jli}^{abc} +8 v_{lij}^{abc} -2 v_{lji}^{abc} \right) \bigg\}
    \end{split}
    \end{equation}
    with
    \begin{equation}
        f_{abc}
            = \left\{
            \begin{split}
                \frac{1}{6}, & \quad{} a = c,   \\
                \frac{1}{2}, & \quad{} a = b~\mathrm{or}~b = c,   \\
                1, & \quad{} \mathrm{otherwise},   \\
            \end{split}
            \right.
    \end{equation}
    \begin{equation}
        w_{ijk}^{abc}
            = W_{ijk}^{abc} / \sqrt{-D_{ijk}^{abc}},
        \quad{}
        v_{ijk}^{abc}
            = V_{ijk}^{abc} / \sqrt{-D_{ijk}^{abc}}
    \end{equation}
    \begin{equation}
        W_{ijk}^{abc}
            = P_{ijk}^{abc} \bigg(
                \sum_{d} V_{bdai} t_{kj}^{cd} -
                \sum_{l} V_{ckjl} t_{il}^{ab}
                % \sum_{d} (bd|ai) t_{kj}^{cd} -
                % \sum_{l} (ck|jl) t_{il}^{ab}
            \bigg)
    \end{equation}
    \begin{equation}
        V_{ijk}^{abc}
            = W_{ijk}^{abc} + \frac{1}{2} P_{ijk}^{abc}
                ( V_{aibj} t_{k}^{c} )
                % [(ai|bj) t_{k}^{c}]
    \end{equation}
    \begin{equation}
        D_{ijk}^{abc}
            = \epsilon_{ii} + \epsilon_{jj} + \epsilon_{kk} -
            \epsilon_{aa} - \epsilon_{bb} - \epsilon_{cc}
    \end{equation}
    where $P_{ijk}^{abc}(\cdot)$ generates 6-fold permutation
    \begin{equation}
        P_{ijk}^{abc} (X_{ijk}^{abc})
            = X_{ijk}^{abc} + X_{ikj}^{acb} + X_{jik}^{bac} + X_{jki}^{bca} + X_{kij}^{cab} + X_{kji}^{cba}
    \end{equation}
    The remaining correlation contribution from orbitals $\notin \mathcal{P}_I$ can be included at $O(N^5)$ cost by invoking a composite correction at MP2 level, leading to the final expression for the LNO-CCSD(T) total correlation energy
    \begin{equation}    \label{eq:Ec_with_dmp2}
        E_{\mathrm{c}}
            = \sum_{I}^{N_{\mathrm{o}}} \left(
                E^{(I)}_{\mathrm{CCSD(T)}} - E^{(I)}_{\mathrm{MP2}}
            \right) + E_{\mathrm{MP2,full}}
    \end{equation}
    where $E^{(I)}_{\mathrm{MP2}}$ is the MP2 correlation energy evaluated in $\mathcal{P}_I$ with $H_I$ and $E_{\mathrm{MP2,full}}$ is the MP2 correlation energy for the entire system.
    The latter can also be decomposed into contributions from each localized orbital
    \begin{equation}    \label{eq:Emp2_decomp}
        E_{\mathrm{MP2,full}}
            = \sum_{I}^{N_{\mathrm{o}}}
            \sum_{jab} \tilde{t}_{Iajb}^{(1)} (2 V_{Iajb} - V_{Ibja})
            = \sum_{I}^{N_{\mathrm{o}}} E^{(I)}_{\mathrm{MP2,full}}
    \end{equation}
    Note that the summation over $jab$ in \cref{eq:Emp2_decomp} is not restricted to $\mathcal{P}_I$ and $\tilde{t}_{Iajb}$ is the exact MP2 amplitude for the entire system with one occupied index being transformed into $\phi_I$, which is not the same as the approximate local MP2 amplitudes defined in \cref{eq:lno_approx_MP2t}.
    Combining \cref{eq:Ec_with_dmp2,eq:Emp2_decomp}, we obtain the final expression of the LNO-CCSD(T) correlation energy
    \begin{equation}    \label{eq:Ec_lnofinal}
        E_{\mathrm{c}}
            = \sum_{I}^{N_{\mathrm{o}}}
            E^{(I)}_{\mathrm{CCSD(T)}} - E^{(I)}_{\mathrm{MP2}} + E^{(I)}_{\mathrm{MP2,full}}
            = \sum_{I}^{N_{\mathrm{o}}} E^{(I)}.
    \end{equation}

    For the purpose of calculating reaction energetics as in this work, the local nature of a chemical reaction allows us to further reduce the computational cost by invoking a strategy resembling the spirit of many fragment embedding methods.
    Specifically, we can choose to use CCSD(T) to evaluate $E^{(I)}$ for only a subset of $M_{\mathrm{o}} \leq N_{\mathrm{o}}$ localized occupied orbitals, with the contributions from the remaining $N_{\mathrm{o}} - M_{\mathrm{o}}$ localized occupied orbitals being accounted for by the MP2 composite correction, leading to a modified energy expression from \cref{eq:Ec_with_dmp2}
    \begin{equation}    \label{eq:Ec_lnoemb}
        E_{\mathrm{c}}
            = \sum_{I}^{M_{\mathrm{o}}}
            (E^{(I)}_{\mathrm{CCSD(T)}} - E^{(I)}_{\mathrm{MP2}})
            + E_{\mathrm{MP2,full}}.
    \end{equation}
    The final reaction energetics calculated from \cref{eq:Ec_lnoemb} thus need to be converged with respect to both the LNO truncation parameters defined in \cref{eq:lno_truncation} and the embedded cluster size $M_{\mathrm{o}}$.
    We detail the convergence of reaction energetics for the two systems studied in this work in \cref{subsec:alo_basis_slab_bz,subsec:tio_basis_slab_bz}, respectively.

    \revision{Summarizing the steps and cost of an LNO calculation, there are three parts.
    \begin{enumerate}
        \item Full-system MP2, which is required by both the LNO construction [Eqs.~(\ref{eq:DIij})--(\ref{eq:lno_occ})] and the MP2 composite correction [Eq.~(\ref{eq:Ec_with_dmp2})] and scales as $O(N^5)$.
        \item $N_{\mathrm{o}}$ independent integral transformations, which are required by the local Hamiltonian construction [Eq.~(\ref{eq:lno_HI})] and scale as $O(N^4 n)$ each, but embarrassingly parallel in $N_{\mathrm{o}}$.
        \item Independent correlated calculations of all local Hamiltonians, which scale as $N_\mathrm{o}$ times the cost of a calculation at the desired level of theory in the local active space, i.e.,~$n^6$ for CCSD and $n^7$ for CCSD(T).
    \end{enumerate}
    For the LNO-CCSD(T) calculations generating the final numbers reported in the main text, solving the local Hamiltonians consists of $90\%$ of the CPU time and hence dominates the computational cost.
    This is due to the relatively large number of LNOs kept in the subspace for fully converging the calculated energetics to an accuracy well within $1$~kcal/mol, as discussed in the main text.
    For calculations of an even larger system or with a looser LNO truncation threshold, the $O(N^5)$ steps may dominate the computational cost.
    Although not explored in this work, many numerical techniques such as local
    domain-based approximations~\cite{Rolik13JCP} and Laplace transform
    methods~\cite{Nagy17JCP} have been exploited to make these steps linear scaling for molecular LNO-CCSD(T) calculations.
    Such advances can also be extended to the periodic LNO-CCSD(T) described here.
    }

    \subsection{Adsorption energy calculation}

    The adsorption energies from PBE0+D3, MP2 and LNO-CCSD(T) are calculated as follows in PySCF
    \begin{equation}
        E_{\textrm{ads}}
            = E_{\textrm{int}}
            + E_{\textrm{surf-relax}}
            + E_{\textrm{mol-relax}}
    \end{equation}
    where
    \begin{equation}
        E_{\textrm{int}}
            = E(\textrm{surf}+\textrm{mol})
            - E(\textrm{surf}+\textrm{ghost-mol})
            - E(\textrm{ghost-surf}+\textrm{mol})
    \end{equation}
    is the adiabatic interaction energy corrected for the basis set superposition error,
    \begin{equation}
        E_{\textrm{surf-relax}}
            = E(\textrm{surf}) - E[\textrm{surf(relaxed)}]
    \end{equation}
    is the surface relaxation energy, and
    \begin{equation}
        E_{\textrm{mol-relax}}
            = E(\textrm{mol}) - E[\textrm{mol(relaxed)}]
    \end{equation}
    is the molecular relaxation energy.
    Both $E_{\textrm{int}}$ and $E_{\textrm{surf-relax}}$ are evaluated with periodic boundary conditions, while $E_{\textrm{mol-relax}}$ is evaluated using the molecular code with the same GTH pseudopotential used for the periodic calculations.

    \subsection{Vibrational corrections}

    \revision{%
    We evaluate the vibrational zero-point energy
    \begin{equation}
        \textrm{ZPE}
            = \sum_{\alpha} \frac{1}{2} \hbar\omega_{\alpha}
    \end{equation}
    and the temperature-dependent vibrational free energy
    \begin{equation}    \label{eq:AvibT}
        A_{\textrm{vib}}(T)
            = -RT \sum_{\alpha} \ln
            \frac{\me^{\beta\hbar\omega_{\alpha}/2}}{\me^{\beta\hbar\omega_{\alpha}} - 1}
    \end{equation}
    using the phonon frequencies $\{\omega_{\alpha}\}$ calculated at the PBE level using the density functional perturbation theory~\cite{Baroni01RMP} (DFPT) as implemented in QE.
    The adsorption energies, reaction energies, and barrier heights reported in Fig.~\fakeref{M2A}, \fakeref{M3D} and \fakeref{M3F} include the ZPE correction.
    The (inverse) reaction rates reported in Fig.~\fakeref{M2B} include the contribution from $A_{\textrm{vib}}(T)$.

    To reduce the high computational cost of the DFPT calculations, we (i) use slab models that are smaller compared to those used in the electronic structure calculations and (ii) freeze a few layers from the bottom, as detailed in \cref{subsec:alo_zpe,subsec:tio_zpe} for the two surface systems.
    The D3 dispersion correction is not included in the DFPT calculations because QE does not support calculating its Hessian.
    However, we verified that all active phonon modes for non-transition state structures and all but one phonon modes for the transition state structures have positive frequencies.
    }

    \clearpage

    \section{Water on the \alosurf{} surface}   \label{sec:alo}

    \subsection{Atomic structure}   \label{subsec:al2o3_atomic_structure}

    \begin{figure}[!h]
        \centering
        \includegraphics[width=1.0\linewidth]{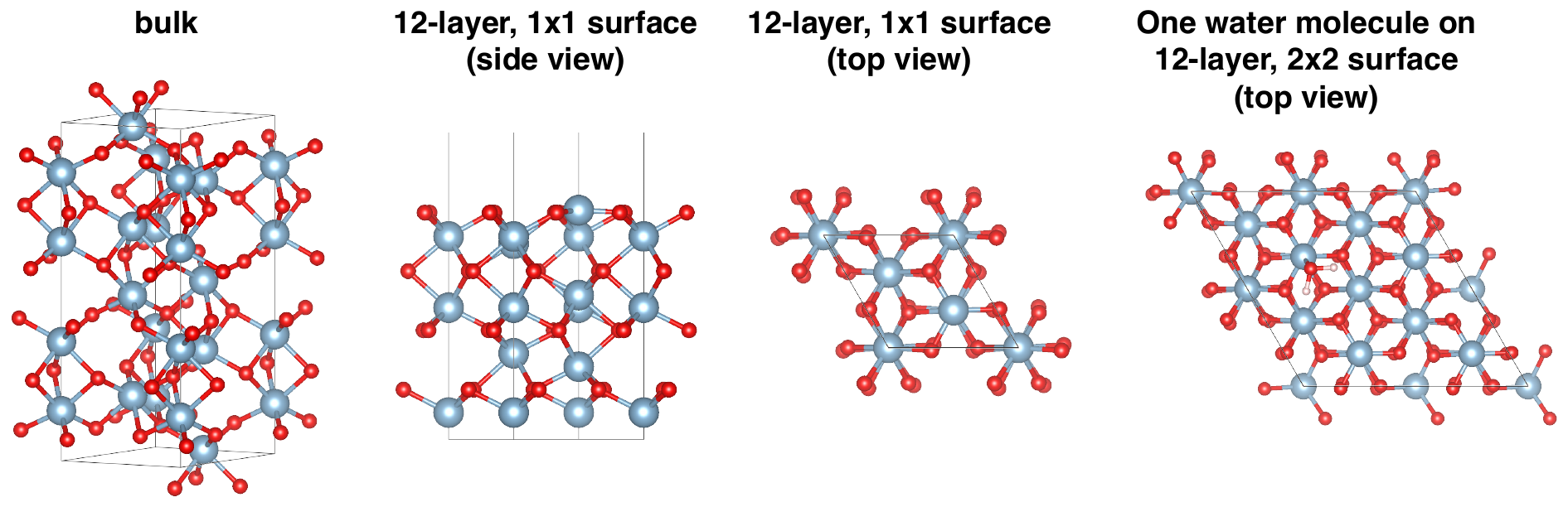}
        \caption{Atomic structure of bulk $\alpha$-\ce{Al2O3}, clean \alosurf{} surface, and water on $2 \times 2$ \alosurf{} surface.}
        \label{fig:al2o3_atomic_structure}
    \end{figure}

    \revision{%
    The atomic structures of bulk $\alpha$-\ce{Al2O3}, clean \alosurf{} surface, and water on $2 \times 2$ \alosurf{} surface are shown in \cref{fig:al2o3_atomic_structure}.
    Following conventions in literature~\cite{Hass98Science,Hass00JPCB,Ranea09JPCC,Wang11JPCC}, each atomic layer is counted as a layer in the slab model.
    This is different from the convention used for the \tiosurf{} surface (see \cref{subsec:tio2_atomic_structure}).
    }

    \subsection{Converged energetics}

    The final numbers for the ZPE-corrected adsorption energy $E_{\textrm{ads}}$, reaction energy $\Delta E$, and barrier $\Delta E^{\ddagger}$ for the water dissociation on \alosurf{} surface reported in Fig.~\fakeref{M2A} are tabulated in \cref{tab:alo_final}.
    These numbers are obtained by combining the energetics from calculations with a finite vacuum size of $23$~$\mathrm{\AA}$ in the direction perpendicular to the slab and a correction to account for an infinite vacuum size ($\Delta_{\infty}$).
    As detailed in the rest of this section, these numbers have been converged to accuracy within $1$~kcal/mol.
    The convergence with respect to basis size, slab/surface size, and Brillouin zone sampling, which gives rise to the numbers listed in column ``$23~\mathrm{\AA}$ vacuum'' in \cref{tab:alo_final}, is detailed in \cref{subsec:alo_basis_slab_bz}.
    The correction to account for an infinite vacuum, which gives rise to the numbers in column ``$\Delta_{\infty}$'', is discussed in \cref{subsec:alo_dipcorr}.
    The ZPE correction evaluated in the harmonic approximation using PBE is discussed in \cref{subsec:alo_zpe}.

    % \begin{landscape}
    % \begin{table}
    \begin{sidewaystable}
        \captionsetup{width=\linewidth}
        \centering
        \caption{Final numbers (highlighted in grey) for the ZPE-corrected adsorption energy $E_{\textrm{ads}}$, reaction energy $\Delta E$, and barrier height $\Delta E^{\ddagger}$ for a single water molecule adsorption and dissociation on the \alosurf{} surface calculated using different methods as reported in Fig.~\fakeref{M2A}.
        Columns labelled by ``$23~\textrm{\AA}$'' list the regular total electronic energy contributions using a supercell with a vacuum of $23~\textrm{\AA}$ in the $z$-direction (see \cref{subsec:alo_basis_slab_bz}).
        Columns labelled by ``$\Delta_{\infty}$'' list the estimated infinite vacuum correction (see \cref{subsec:alo_dipcorr}).
        Columns labelled by ``ZPE'' list the zero-point energy correction (see \cref{subsec:alo_zpe}).
        All numbers are in kcal/mol.}
        \label{tab:alo_final}
        \begin{tabular}{lcccacccaccca}
            \hline\hline
            \multirow{2}*{Method}
                & \multicolumn{4}{c}{$E_{\textrm{ads}}$}
                & \multicolumn{4}{c}{$\Delta E$}
                & \multicolumn{4}{c}{$\Delta E^{\ddagger}$}   \\
            \cmidrule(lr){2-5} \cmidrule(lr){6-9} \cmidrule(lr){10-13}
                & 23 $\mathrm{\AA}$ & $\Delta_{\infty}$ & ZPE & Final
                & 23 $\mathrm{\AA}$ & $\Delta_{\infty}$ & ZPE & Final
                & 23 $\mathrm{\AA}$ & $\Delta_{\infty}$ & ZPE & Final    \\
            \hline
            % PBE+D3      & $-29.2$ & $-0.5$ & $0.4$ & $-29.3$
            %             & $-9.3$ & $1.4$ & $-0.6$ & $-8.5$
            %             & $4.5$  & $+0.5$ & $-2.2$ & $2.8$   \\
            % PBE0+D3     & $-30.1$ & $-0.5$ & $0.4$ & $-30.2$
            %             & $-9.2$ & $1.4$ & $-0.6$ & $-8.4$
            %             & $6.2$  & $+0.5$ & $-2.2$ & $4.5$   \\
            % SCAN+D3     & $-30.4$ & $-0.5$ & $0.4$ & $-30.5$
            %             & $-8.9$ & $1.4$ & $-0.6$ & $-8.2$
            %             & $5.7$  & $+0.5$ & $-2.2$ & $4.0$   \\
            % & & & & & & & & & & & & \\
            % MP2         & $-27.3$ & $0.0$ & $0.4$ & $-26.9$
            %             & $-9.1$ & $1.8$ & $-0.6$ & $-7.9$
            %             & $7.0$  & $+0.7$ & $-2.2$ & $5.5$   \\
            % LNO-CCSD(T) & $-27.6$ & $0.0$ & $0.4$ & $-27.2$
            %             & $-8.9$ & $1.8$ & $-0.6$ & $-7.7$
            %             & $8.1$  & $+0.7$ & $-2.2$ & $6.6$   \\
            PBE+D3 & $-29.21$ & $-0.62$ & $0.42$ & $-29.41$ & $-9.29$ & $1.35$ & $-0.62$ & $-8.56$ & $4.51$ & $0.52$ & $-2.21$ & $2.82$ \\
            PBE0+D3 & $-29.87$ & $-0.62$ & $0.42$ & $-30.07$ & $-9.17$ & $1.35$ & $-0.62$ & $-8.44$ & $6.22$ & $0.52$ & $-2.21$ & $4.53$ \\
            SCAN+D3 & $-30.40$ & $-0.62$ & $0.42$ & $-30.60$ & $-8.94$ & $1.35$ & $-0.62$ & $-8.21$ & $5.68$ & $0.52$ & $-2.21$ & $3.99$ \\
            & & & & & & & & & & & &  \\
            MP2 & $-26.03$ & $-0.74$ & $0.42$ & $-26.35$ & $-9.07$ & $2.05$ & $-0.62$ & $-7.64$ & $7.03$ & $0.75$ & $-2.21$ & $5.57$ \\
            CCSD(T) & $-26.22$ & $-0.74$ & $0.42$ & $-26.54$ & $-8.93$ & $2.05$ & $-0.62$ & $-7.50$ & $8.07$ & $0.75$ & $-2.21$ & $6.61$    \\
            \hline
        \end{tabular}
    \end{sidewaystable}
    % \end{table}
    % \end{landscape}

    \subsection{Slab size, surface size, basis size, and Brillouin zone sampling}
    \label{subsec:alo_basis_slab_bz}

    We follow Ref.~\citenum{Hass98Science} and other previous DFT studies (e.g., Refs.~\citenum{Hass00JPCB,Ranea09JPCC,Wang11JPCC,Wirth12JPCC}) of the same system for the notation of slab size (in terms of the number of atomic layers) and surface size.
    We will use a notation $(n,m)$ for a slab of $n$ layers, with the top $m$ layers being allowed to relax during geometry optimization.
    A vacuum of $23$~$\mathrm{\AA}$ perpendicular to the slab plane is used throughout this section.
    The determination of a correction to account for the infinite vacuum limit is detailed in \cref{subsec:alo_dipcorr}.

    \noindent
    \textbf{PBE+D3 and SCAN+D3}.
    By first fixing the slab size to be $(12,8)$, we determined from \cref{tab:alo_dft_kesurfkpt} the following converged parameters for the plane wave PBE+D3 calculations:
    \begin{itemize}
        \item a kinetic energy cutoff of $50$~Hartree or $1350$~eV,
        \item a surface size of $2 \times 2$,
        \item Brillouin zone sampling at the $\Gamma$ point.
    \end{itemize}
    By fixing these parameters and scanning over different slab sizes, we determined from \cref{tab:alo_dft_slab} a converged slab size of $(12,8)$.
    These parameters are also used for the SCAN+D3 calculations, except that we use a higher kinetic energy cutoff of $2000$~eV due to the harder pseudopotential.

    \begin{table}[!hbtp]
        \centering
        \caption{PBE+D3 dissociation energy $\Delta E_{\tr{diss}}$ and the barrier height $\Delta E^{\ddagger}$ (both in kcal/mol) of a single adsorbed water on the \alosurf{} surface evaluated for a $(12,8)$ slab model with different kinetic energy cutoff (in Hartree), surface size, and Brillouin zone sampling.
        The parameters that converge both energies within $1$~kcal/mol are highlighted in grey.}
        \label{tab:alo_dft_kesurfkpt}
        \begin{tabular}{lllll}
            \hline\hline
            K.E.~cutoff & surface & $k$-point mesh & $\Delta E_{\tr{diss}}$ & $\Delta E^{\ddagger}$  \\
            \hline
            \rowcolor{Gray}
            $50$ & $2\times{}2$ & $1\times{}1\times{}1$ & $-9.3$ & $4.6$    \\
            $100$ & $2\times{}2$ & $1\times{}1\times{}1$ & $-9.3$ & $4.6$    \\
            $100$ & $2\times{}2$ & $2\times{}2\times{}1$ & $-9.3$ & $4.6$    \\
            $50$ & $3\times{}3$ & $1\times{}1\times{}1$ & $-9.0$ & $4.6$    \\
            \hline
        \end{tabular}
    \end{table}

    \begin{table}[!hbtp]
        \centering
        \caption{Same reaction energetics as in \cref{tab:alo_dft_kesurfkpt} for different choices of slab size.
        All other parameters are fixed at their optimum values determined in \cref{tab:alo_dft_kesurfkpt}.
        The slab size that converge both energies within $1$~kcal/mol is highlighted in grey.}
        \label{tab:alo_dft_slab}
        \begin{tabular}{llll}
            \hline\hline
            Slab size & $\Delta E_{\tr{diss}}$ & $\Delta E^{\ddagger}$ \\
            \hline
            (12,5) & $-9.0$ & $4.4$    \\
            (12,7) & $-9.2$ & $4.6$    \\
            \rowcolor{Gray}
            (12,8) & $-9.3$ & $4.6$    \\
            (15,10) & $-9.4$ & $4.6$    \\
            \hline
        \end{tabular}
    \end{table}

    \noindent
    \textbf{PBE0+D3 and MP2}.
    By fixing the slab size to be $(9,5)$, we determined from \cref{tab:alo_hfmp2_basis} that a TZ Gaussian basis set is sufficient for converged energetics for both mean-field (HF and PBE) and correlated (MP2) calculations.
    The same table also confirms that the PBE+D3 energetics from PySCF with the TZ Gaussian basis set agrees with the plane wave PBE+D3 numbers from Quantum Espresso (the remaining difference, which is less than $1$~kcal/mol, is mostly caused by the use of different pseudopotentials).
    By fixing the basis set and scanning over different slab sizes, we verified with data in \cref{tab:alo_mp2_slab} that the $(12,8)$ slab, which converges the PBE+D3 energetics as concluded from \cref{tab:alo_dft_slab}, is also sufficient to converge the MP2 energetics.
    Finally, we also expect these parameters (TZ basis set and $(12,8)$ slab) to be sufficient for PBE0+D3 calculations, as the PBE0 energy is a mixture of the PBE energy and the HF exchange energy.

    \begin{table}[!hbtp]
        \centering
        \caption{Reaction energetics (in kcal/mol) from HF, MP2, and PBE+D3 calculated using PySCF with the GTH-cc-pV$X$Z basis sets for a $(9,5)$/$2 \times 2$ surface model with $\Gamma$-point Brillouin zone sampling.
        A TZ basis set (highlighted in grey) is sufficient to converge reaction energetics to within $1$~kcal/mol for both mean-field (HF, PBE, and hence also PBE0) and correlated (MP2) calculations.
        The basis-set converged PBE+D3 reaction energetics also agree with those from plane wave/Quantum Espresso very well (the small difference is likely caused by the difference in pseudopotentials).
        }
        \label{tab:alo_hfmp2_basis}
        \begin{tabular}{lllllll}
            \hline\hline
            \multirow{2}*{Basis set}
                & \multicolumn{2}{c}{HF}
                & \multicolumn{2}{c}{MP2}
                & \multicolumn{2}{c}{PBE}   \\
            \cmidrule(lr){2-3}
            \cmidrule(lr){4-5}
            \cmidrule(lr){6-7}
                & $\Delta E_{\tr{diss}}$ & $\Delta E^{\ddagger}$
                & $\Delta E_{\tr{diss}}$ & $\Delta E^{\ddagger}$
                & $\Delta E_{\tr{diss}}$ & $\Delta E^{\ddagger}$   \\
            \hline
            DZ/PySCF & $-13.6$ & $12.6$ & $-8.4$ & $7.2$ & $-7.6$ & $5.0$ \\
            \rowcolor{Gray}
            TZ/PySCF & $-13.6$ & $13.5$ & $-8.7$ & $7.2$ & $-8.7$ & $5.3$ \\
            QZ/PySCF & $-13.6$ & $13.5$ & $-8.7$ & $7.2$ & $-8.7$ & $5.3$ \\
            \hline
            PW/QE &      &        &        &       & $-9.0$ & $4.7$ \\
            \hline
        \end{tabular}
    \end{table}

    \begin{table}[!hbtp]
        \centering
        \caption{Reaction energetics (in kcal/mol) from MP2 calculated for different slab models using a TZ basis set and for a $2 \times 2$ surface model with $\Gamma$-point Brillouin zone sampling.
        As in \cref{tab:alo_dft_slab}, the same $(12,8)$ slab model is sufficient to converge the MP2 energy.}
        \label{tab:alo_mp2_slab}
        \begin{tabular}{lll}
            \hline\hline
                Slab & $\Delta E_{\tr{diss}}$ & $\Delta E^{\ddagger}$  \\
            \hline
                (12,5) & $-9.1$ & $7.1$    \\
                (12,7) & $-9.0$ & $7.2$    \\
                \rowcolor{Gray}
                (12,8) & $-9.1$ & $7.0$    \\
                (15,10) & $-9.1$ & $7.4$    \\
            \hline
        \end{tabular}
    \end{table}

    \noindent
    \textbf{LNO-CCSD(T)}.
    As detailed in \cref{subsec:lnocc_theory}, we use \cref{eq:Ec_lnoemb} to evaluate the LNO-CCSD(T) reaction energetics, which must be converged with respect to both the LNO truncation parameter, $\lambda_{\mathrm{o}}$ and $\lambda_{\mathrm{v}}$, defined in \cref{eq:lno_truncation},
    and the ``embedded'' cluster size, $M_{\mathrm{o}}$, defined in \cref{eq:Ec_lnoemb}.
    For the LNO truncation parameters, we follow Ref.~\citenum{Rolik13JCP} and scan $\lambda_{\mathrm{v}}$ while keeping $\lambda_{\mathrm{o}} = 10 \lambda_{\mathrm{v}}$.
    For the embedded cluster size $M_{\mathrm{o}}$, we note that the localized occupied orbitals (LOs) for this system fall into two classes: $4$ LOs on the water molecule and $48$ LOs for every layer of oxygen atoms (there are no LOs localized on aluminum due to the electron negativity difference between aluminum and oxygen).
    This results in a natural way of systematically increasing $M_{\mathrm{o}}$:
    \begin{enumerate}
        \item LOs on the water molecule only, $M_{\mathrm{o}} = 4$,
        \item plus LOs on the first oxygen layer, $M_{\mathrm{o}} = 4 + 48 = 52$,
        \item plus LOs on the second oxygen layer, $M_{\mathrm{o}} = 4 + 48 + 48 = 100$,
        \item plus LOs on the third oxygen layer, $M_{\mathrm{o}} = 4 + 48 + 48 + 48 = 148$,
        \item plus LOs on the third oxygen layer, $M_{\mathrm{o}} = 4 + 48 + 48 + 48 + 48 = 196 = N_{\mathrm{o}}$.
    \end{enumerate}

\begin{figure}[!htbp]
        \centering
        \includegraphics[width=1.0\linewidth]{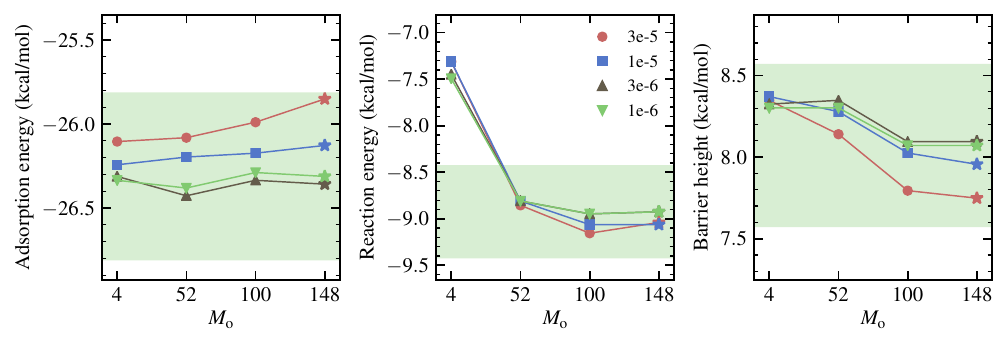}
        \captionsetup{width=\linewidth}
        \caption{\revision{Convergence of the LNO-CCSD(T) adsorption energy (left), reaction energy (middle), and barrier height (right) for water-\alo{} with respect to the embedded cluster size $M_{\mathrm{o}}$.
        A $(12,8)$/$2 \times 2$ surface model and a TZ basis set are employed.
        Data of different color correspond to different virtual LNO truncation parameter $\lambda_{\mathrm{v}}$ (the occupied truncation parameter is chosen to be $\lambda_{\mathrm{o}} = 10\lambda_{\mathrm{v}}$ in all cases).
        The range of $\pm 0.5$~kcal/mol from the most converged number (i.e.,~$\lambda_{\textrm{v}} = 10^{-6}$ and $M_{\textrm{o}} = 148$ is highlighted by the green shaded area.)
        }
        }
        \label{fig:alo_lnoconv}
    \end{figure}

    The convergence with respect to $M_{\mathrm{o}}$ for different choices of $\lambda_{\mathrm{v}}$ is shown in \cref{fig:alo_lnoconv} for a $(12,8)$/$2 \times 2$ surface model using a TZ basis set.
    \revision{For both the adsorption energy and the barrier height, convergence to an accuracy better than $1$~kcal/mol is achieved using even the $4$ LOs on the water molecule, while the reaction energy shows a slower convergence and requires further including the first oxygen layer (i.e.,~$M_{\textrm{o}} = 52$).
    Nevertheless, for the numbers reported in \cref{tab:alo_final} and in the main text, we use the fully converged results from $M_{\textrm{o}} = 148$.
    }
    % including LOs up to the third oxygen layer with $M_{\mathrm{o}} = 148$ is sufficient to converge the reaction energetics to accuracy within $1$~kcal/mol.
    The $M_{\mathrm{o}}$-converged reaction energies and barrier heights from different choices of $\lambda_{\mathrm{v}}$ (marked as stars in \cref{fig:alo_lnoconv}) are then plotted in Fig.~\fakeref{M1E} to investigate the convergence with respect to the LNO truncation, where we also see a fast convergence to the desired accuracy of $1$~kcal/mol, as already discussed in the main text.
    The convergence with respect to surface size, slab thickness, and basis size is presented in Fig.~\fakeref{M1G} and \fakeref{M1H}, from which we determine that the $(12,8)$/$2 \times 2$ model and the TZ basis set are sufficient for converging both the reaction energy and the barrier height to chemical accuracy.
    \revision{These parameters are then used to evaluate the adsorption energy without further modifications.}

    \subsection{Infinite vacuum correction}
    \label{subsec:alo_dipcorr}

    \noindent
    \textbf{PBE+D3, SCAN+D3, and PBE0+D3}.
    We use the dipole correction at PBE+D3 level as implemented in Quantum Espresso to account for the finite-vacuum effect.
    For the $(12,8)$ slab model, we found $\Delta_{\infty} = -0.62$, $+1.35$, and $+0.52$~kcal/mol for the adsorption energy, reaction energy, and barrier height, respectively, as also tabulated in \cref{tab:alo_final}.
    These corrections are also applied to obtain the final SCAN+D3 and PBE0+D3 energetics in \cref{tab:alo_final}.

\begin{figure}[!htbp]
        \centering
        \includegraphics[width=1.0\linewidth]{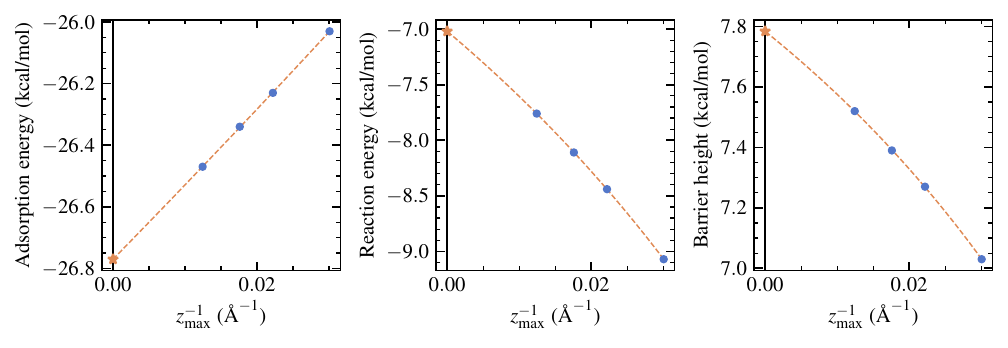}
        \caption{Convergence of the MP2 adsorption energy (left), reaction energy (middle), and barrier height (right) for water-\alo{} as a function of the reciprocal dimension perpendicular to the surface.
        Extrapolations based on \cref{eq:alo_vacextrap} are shown in orange.
        }
        \label{fig:alo_mp2_vacconv}
    \end{figure}

    \noindent
    \textbf{MP2 and LNO-CCSD(T)}.
    For MP2, we manually increase the dimension perpendicular to the surface ($z_{\textrm{max}}$) and repeat the calculations using the $(12,8)$ model and the TZ basis set.
    \revision{The resulting energies are plotted in \cref{fig:alo_mp2_vacconv}, where extrapolation to the infinite vacuum limit using the following function is also shown
    \begin{equation}    \label{eq:alo_vacextrap}
        E(z_{\textrm{max}})
            = \frac{A}{z_{\mathrm{max}} - B} + E(z_{\mathrm{max}} = \infty)
    \end{equation}
    From the plot, we obtain $\Delta_{\infty} = -0.74$, $2.05$, and $0.75$~kcal/mol for $E_{\textrm{ads}}$, $\Delta E$, and $E^{\ddagger}$, respectively, as tabulated in \cref{tab:alo_final}.
    }
    In the spirit of the MP2 composite correction, these corrections are also applied to obtain the final LNO-CCSD(T) energetics in \cref{tab:alo_final}.

    \subsection{Vibrational corrections}
    \label{subsec:alo_zpe}

\begin{figure}
        \centering
        \includegraphics[width=0.5\linewidth]{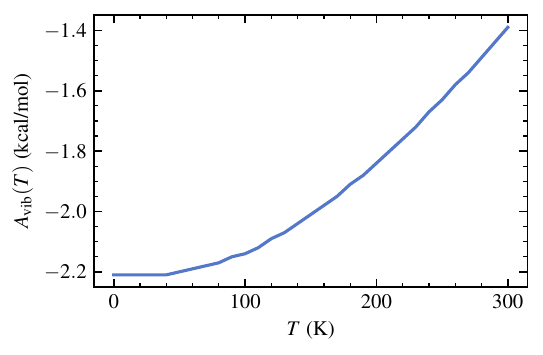}
        \caption{The temperature-dependent vibrational free energy for the barrier height of water dissociation on \alo{} surface calculated according to \cref{eq:AvibT}.}
        \label{fig:alo_AvibT}
    \end{figure}

    \revision{%
    DFPT calculations at PBE level are performed for the clean surface, molecular adsorption, dissociated adsorption, and transition state of a (9,5)/$2 \times 2$ surface model.
    Atoms in the four layers from the bottom are kept frozen in the DFPT calculations to reduce the computational cost.
    The reciprocal space is sampled at the $\Gamma$-point.
    The ZPE corrections for the adsorption energy, reaction energy, and barrier height are already listed in \cref{tab:alo_final}.
    The temperature-dependent vibrational activation free energy used to evaluate the reaction rates shown in Fig.~\fakeref{M2B} is plotted in \cref{fig:alo_AvibT}.
    }

    \clearpage

    \section{Water on the \tiosurf{} surface}   \label{sec:tio}

    \subsection{Atomic structure}   \label{subsec:tio2_atomic_structure}

    \begin{figure}[!h]
        \centering
        \includegraphics[width=1.0\linewidth]{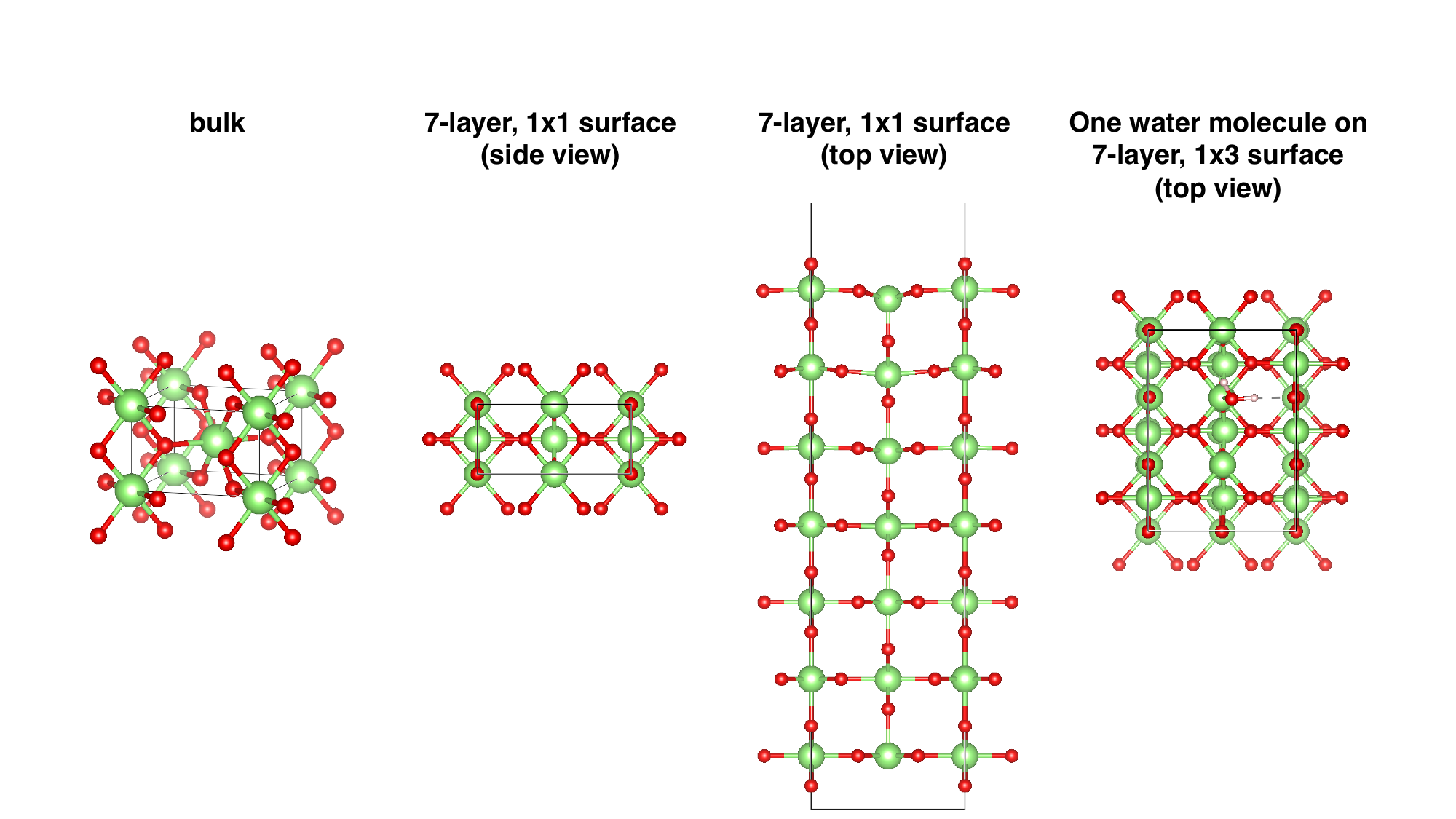}
        \caption{Atomic structure of bulk rutile \ce{TiO2}, clean \tiosurf{} surface, and water on $1 \times 3$ \tiosurf{} surface.}
        \label{fig:tio2_atomic_structure}
    \end{figure}

    \revision{%
    The atomic structures of bulk rutile \ce{TiO2}, clean \tiosurf{} surface, and water on $1 \times 3$ \tiosurf{} surface are shown in \cref{fig:tio2_atomic_structure}.
    The definition of atomic layer follows the literature~\cite{Harris04PRL,Lindan05PRB,Wang17PNAS}, where an atomic layer includes a [Ti-O] plane plus the two O planes above and beneath it.
    Note that this is different from the convention used in \cref{subsec:al2o3_atomic_structure} for water on \alosurf{} surface.
    }

    \subsection{Converged energetics}

    The final numbers for the ZPE-corrected adsorption energy $E_{\textrm{ads}}$, reaction energy $\Delta E$, and barrier height $\Delta E^{\ddagger}$ for the water adsorption and dissociation on \tiosurf{} surface reported in Fig.~\fakeref{M3D} and \fakeref{M3F} are tabulated in \cref{tab:tio_final}.
    These numbers are obtained by combining the energetics from calculations with a finite vacuum size of $18$~$\mathrm{\AA}$ in the direction perpendicular to the slab and a correction to account for an infinite vacuum size ($\Delta_{\infty}$).
    As detailed in the rest of this section, these numbers have been converged to accuracy within $1$~kcal/mol.
    The convergence with respect to basis size, slab/surface size, and Brillouin zone sampling, which gives rise to the numbers listed in column ``$18~\mathrm{\AA}$ vacuum'' in \cref{tab:tio_final}, is detailed in \cref{subsec:tio_basis_slab_bz}.
    The correction to account for an infinite vacuum, which gives rise to the numbers in column ``$\Delta_{\infty}$'', is discussed in \cref{subsec:tio_dipcorr}.
    The ZPE correction evaluated in the harmonic approximation using PBE is discussed in \cref{subsec:tio_zpe}.

    % \begin{landscape}
    %     \global\pdfpageattr\expandafter{\the\pdfpageattr/Rotate 90}
    % \begin{table}[!h]
    \begin{sidewaystable}
        \captionsetup{width=\linewidth}
        \centering
        \caption{Final numbers (highlighted in grey) for the ZPE-corrected adsorption energy $E_{\textrm{ads}}$, reaction energy $\Delta E$, and barrier height $\Delta E^{\ddagger}$ for a single water molecule adsorption and dissociation on the \tiosurf{} surface calculated using different methods as reported in Fig.~\fakeref{M3D} and \fakeref{M3F}.
        Columns labelled by ``$18~\textrm{\AA}$'' list the regular total electronic energy contributions using a supercell with a vacuum of $18~\textrm{\AA}$ in the $z$-direction (see \cref{subsec:tio_basis_slab_bz}).
        Columns labelled by ``$\Delta_{\infty}$'' list the estimated infinite vacuum correction (see \cref{subsec:tio_dipcorr}).
        Columns labelled by ``ZPE'' list the zero-point energy correction (see \cref{subsec:tio_zpe}).
        All numbers are in kcal/mol.}
        \label{tab:tio_final}
        \begin{tabular}{lcccacccaccca}
            \hline\hline
            \multirow{2}*{Method}
                & \multicolumn{4}{c}{$E_{\textrm{ads}}$}
                & \multicolumn{4}{c}{$\Delta E$}
                & \multicolumn{4}{c}{$\Delta E^{\ddagger}$}   \\
            \cmidrule(lr){2-5} \cmidrule(lr){6-9} \cmidrule(lr){10-13}
                & 18 $\mathrm{\AA}$ & $\Delta_{\infty}$ & ZPE & Final
                & 18 $\mathrm{\AA}$ & $\Delta_{\infty}$ & ZPE & Final
                & 18 $\mathrm{\AA}$ & $\Delta_{\infty}$ & ZPE & Final    \\
            \hline
            PBE+D3 & $-22.63$ & $0.20$ & $1.40$ & $-21.03$ & $0.65$ & $-0.03$ & $-1.55$ & $-0.93$ & $6.26$ & $-0.09$ & $-2.82$ & $3.35$ \\
            PBE+U(1)+D3 & $-21.34$ & $0.20$ & $1.40$ & $-19.74$ & $0.42$ & $-0.03$ & $-1.55$ & $-1.16$ & $5.84$ & $-0.09$ & $-2.82$ & $2.93$ \\
            PBE+U(3)+D3 & $-18.60$ & $0.20$ & $1.40$ & $-17.00$ & $-0.16$ & $-0.03$ & $-1.55$ & $-1.74$ & $4.87$ & $-0.09$ & $-2.82$ & $1.96$ \\
            PBE+U(5)+D3 & $-15.64$ & $0.20$ & $1.40$ & $-14.04$ & $-0.91$ & $-0.03$ & $-1.55$ & $-2.49$ & $3.69$ & $-0.09$ & $-2.82$ & $0.78$ \\
            PBE+U(8)+D3 & $-10.88$ & $0.20$ & $1.40$ & $-9.28$ & $-2.42$ & $-0.03$ & $-1.55$ & $-4.00$ & $1.51$ & $-0.09$ & $-2.82$ & $-1.40$ \\
            PBE+U(11)+D3 & $-5.90$ & $0.20$ & $1.40$ & $-4.30$ & $-4.30$ & $-0.03$ & $-1.55$ & $-5.88$ & $-1.06$ & $-0.09$ & $-2.82$ & $-3.97$ \\
            SCAN+D3 & $-25.91$ & $0.20$ & $1.40$ & $-24.31$ & $1.31$ & $-0.03$ & $-1.55$ & $-0.27$ & $7.45$ & $-0.09$ & $-2.82$ & $4.54$ \\
            & & & & & & & & & & & &  \\
            MP2 & $-25.65$ & $0.51$ & $1.40$ & $-23.74$ & $0.92$ & $0.07$ & $-1.55$ & $-0.56$ & $9.28$ & $-0.15$ & $-2.82$ & $6.31$ \\
            CCSD(T) & $-22.74$ & $0.51$ & $1.40$ & $-20.83$ & $-0.41$ & $0.07$ & $-1.55$ & $-1.89$ & $7.06$ & $-0.15$ & $-2.82$ & $4.09$ \\
            \hline
        \end{tabular}
    \end{sidewaystable}

    \subsection{Slab size, surface size, basis size, and Brillouin zone sampling}
    \label{subsec:tio_basis_slab_bz}

    We follow Ref.~\citenum{Lindan05PRB} and other DFT studies on the same system for the notation of slab size and surface size.
    Like in water-\alo{}, we use a notation $(n,m)$ to denote a slab of $n$ layers, with the top $m$ layers being allowed to relax during geometry optimization

\begin{figure}[!htbp]
        \centering
        \includegraphics[width=0.3\linewidth]{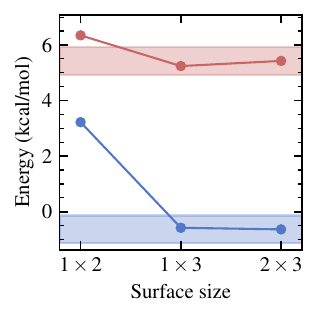}
        \hspace{1em}
        \includegraphics[width=0.3\linewidth]{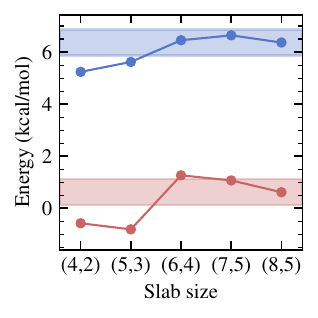}
        \hspace{1em}
        \includegraphics[width=0.3\linewidth]{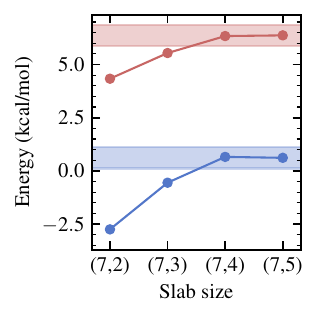}
        \captionsetup{width=\linewidth}
        \caption{Convergence of the PBE+D3 reaction energy (blue) and barrier height (red) for water dissociation on \tio{} surface with respect to the surface size (left), the total slab size (middle), and the active slab size (right).
        A $(4,2)$ slab model is used for the left panel, while a $1 \times 3$ surface model is used for the other two.
        All calculations were done with the Brillouin zone sampled by a $2\times 2 \times 1$ mesh.}
        \label{fig:tio_pbe_surflayer}
    \end{figure}

    \noindent
    \textbf{PBE+D3, PBE+$U$+D3, and SCAN+D3}.
    Our preliminary tests suggest that a kinetic energy cutoff of $1350$~eV and a $2\times 2\times 1$ $k$-point mesh properly converge the basis set size and the Brillouin zone for the PBE+D3 reaction energetics.
    By fixing these parameters, we determined a converged surface size of $1 \times 3$ and a converged slab model of $(7,4)$ from \cref{fig:tio_pbe_surflayer}.
    The same parameters are used for PBE$+U$+D3 and SCAN+D3, except that we use a higher kinetic energy cutoff of $2000$~eV for SCAN+D3 due to the harder pseudopotential.

    \noindent
    \textbf{MP2}.
    \Cref{tab:tio_mp2_basis} suggests that a TZ basis set is sufficient to converge reaction energetics evaluated at both mean-field (HF) and correlated (MP2) level.
    \Cref{tab:tio_mp2_kpt} suggests that the HF part of the MP2 energy requires sampling the Brillouin zone with a $2 \times 2 \times 1$ mesh, while the correlation part converges faster and requires only $\Gamma$-point Brillouin sampling.
    Ideally, we thus would like to use the following energy expression for the total MP2 energy
    \begin{equation}    \label{eq:tio_mp2}
        E^{(7,4)}_{\mathrm{MP2}}(\mathrm{TZ}/2\times 2\times 1)
            \approx E^{(7,4)}_{\mathrm{HF}}(\mathrm{TZ}/2\times 2\times 1)
            + E^{(7,4)}_{\mathrm{MP2,c}}(\mathrm{TZ}/1\times 1\times 1)
    \end{equation}
    However, neither the TZ/$2\times 2\times 1$ HF calculation nor the TZ/$1\times 1\times 1$ MP2 calculation is feasible with the available computational resources.
    We thus approximate them using the following composite correction
    \begin{equation}    \label{eq:tio_hf_compcorr}
    \begin{split}
        E^{(7,4)}_{\mathrm{HF}}(\mathrm{TZ}/2\times 2\times 1)
            \approx &~E^{(7,4)}_{\mathrm{HF}}(\mathrm{DZ}/1\times 1\times 1)  \\
        \phantom{E^{(7,4)}_{\mathrm{HF}}(\mathrm{TZ}/2\times 2\times 1)\approx}&
            + \underbrace{
                E^{(6,4)^*}_{\mathrm{HF}}(\mathrm{TZ}/1\times 1\times 1)
                - E^{(6,4)^*}_{\mathrm{HF}}(\mathrm{DZ}/1\times 1\times 1)
            }_{\Delta E^{(6,4)^*}_{\mathrm{HF}}(\mathrm{TZ}/1\times 1\times 1)}   \\
        \phantom{E^{(7,4)}_{\mathrm{HF}}(\mathrm{TZ}/2\times 2\times 1)\approx}&
            + \underbrace{
                E^{(6,4)^*}_{\mathrm{HF}}(\mathrm{DZ}/2\times 2\times 1)
                - E^{(6,4)^*}_{\mathrm{HF}}(\mathrm{DZ}/1\times 1\times 1)
            }_{\Delta E^{(6,4)^*}_{\mathrm{HF}}(\mathrm{DZ}/2\times 2\times 1)}
    \end{split}
    \end{equation}
    \begin{equation}    \label{eq:tio_mp2c_compcorr}
    \begin{split}
        E^{(7,4)}_{\mathrm{MP2,c}}(\mathrm{TZ}/1\times 1\times 1)
            \approx&~E^{(7,4)}_{\mathrm{MP2,c}}(\mathrm{DZ}/1\times 1\times 1)   \\
        \phantom{E^{(7,4)}_{\mathrm{MP2,c}}(\mathrm{TZ}/1\times 1\times 1)\approx}&
            + \underbrace{
                E^{(6,4)^*}_{\mathrm{MP2,c}}(\mathrm{TZ}/1\times 1\times 1)
                - E^{(6,4)^*}_{\mathrm{MP2,c}}(\mathrm{DZ}/1\times 1\times 1)
            }_{\Delta E^{(6,4)^*}_{\mathrm{MP2,c}}(\mathrm{TZ}/1\times 1\times 1)}
    \end{split}
    \end{equation}
    where $(6,4)^*$ is derived from the (7,4) slab model by removing one atomic layer from the bottom (this layer was kept frozen during the geometry relaxation).
    The final MP2 reaction energy and barrier height estimated using \cref{eq:tio_mp2,eq:tio_hf_compcorr,eq:tio_mp2c_compcorr} are shown in \cref{tab:tio_mp2_compcorr}, along with all the energy components.
    % However, a TZ/$2\times 2\times 1$ HF calculation for the large $(7.4)$ slab model we want to study is infeasible with the available computational resources.
    % We thus designed the following strategy
    % \begin{equation}    \label{eq:tio_mp2}
    % \begin{split}
    %     E_{\mathrm{MP2}}(\mathrm{TZ}/2\times 2\times 1)
    %         &\approx E_{\mathrm{MP2,c}}(\mathrm{TZ}/1\times 1\times 1) +
    %         E_{\mathrm{HF}}(\mathrm{DZ}/2\times 2\times 1) + \\
    %         &\quad{} E_{\mathrm{HF}}(\mathrm{TZ}/1\times 1\times 1) -
    %         E_{\mathrm{HF}}(\mathrm{DZ}/1\times 1\times 1)
    % \end{split}
    % \end{equation}
    % where the desired TZ/$2 \times 2 \times 1$ HF energy is approximated in a composite-correction way.

    \begin{table}[!htbp]
        \centering
        \caption{Reaction energy $\Delta E$ and barrier $\Delta E^{\ddagger}$ of water dissociation on \tio{} for a $(4,2)$ slab model with $1 \times 2$ surface calculated using MP2 with $\Gamma$-point Brillouin zone sampling and different Gaussian basis sets.
        Contributions from the HF and the MP2 correlation energy components to the MP2 energetics are also shown.
        As highlighted in gray, both the HF energy and the MP2 correlation energy contributions to the reaction energetics converge to accuracy within $1$~kcal/mol with a TZ basis set.}
        \label{tab:tio_mp2_basis}
        \begin{tabular}{ccccccc}
            \hline\hline
            \multirow{2}*{Basis set}
                & \multicolumn{3}{c}{$\Delta E$}
                & \multicolumn{3}{c}{$\Delta E^{\ddagger}$}    \\
            \cmidrule(lr){2-4} \cmidrule(lr){5-7}
                & HF & MP2,c & MP2
                & HF & MP2,c & MP2  \\
            \hline
            DZ & $0.5$     & $6.1$ & $6.6$
               & $10.8$    & $0.4$ & $11.2$    \\
            \CG{}TZ & \CG$-1.1$ & \CG$7.0$    & \CG$5.9$
               & \CG$9.9$  & \CG$-0.2$   & \CG$9.7$     \\
            QZ & $-1.4$    & $7.3$    & $5.9$
               & $9.7$     & $0.4$    & $10.1$    \\
            \hline
        \end{tabular}
    \end{table}

    \begin{table}[!htbp]
        \centering
        \caption{Same reaction energetics as in \cref{tab:tio_mp2_basis} but calculated using MP2 with a DZ basis set and different Brillouin zone sampling.
        Contributions from the HF and the MP2 correlation energy components to the MP2 energetics are also shown.
        As highlighted in gray, the HF energy converges to accuracy within $1$~kcal/mol with $2 \times 2 \times 1$ Brillouin zone sampling, while the MP2 correlation energy converges even faster with $\Gamma$-point Brillouin zone sampling.}
        \label{tab:tio_mp2_kpt}
        \begin{tabular}{ccccccc}
            \hline\hline
            \multirow{2}*{BZ sampling}
                & \multicolumn{3}{c}{$\Delta E$}
                & \multicolumn{3}{c}{$\Delta E^{\ddagger}$}    \\
            \cmidrule(lr){2-4} \cmidrule(lr){5-7}
                & HF & MP2,c & MP2
                & HF & MP2,c & MP2  \\
            \hline
            $1\times 1 \times 1$ & $0.5$     & \CG$6.1$ & $6.6$
                & $10.8$ & \CG$0.4$  & $11.2$    \\
            $2\times 2 \times 1$ & \CG$-1.5$ & $6.2$    & $4.7$
                & \CG$10.7$ & $-0.1$ & $10.6$    \\
            $3\times 3 \times 1$ & $-1.7$    &          &
                & $10.7$ &        &           \\
            \hline
        \end{tabular}
    \end{table}

    \begin{table}[!htbp]
        \centering
        \caption{Estimating the MP2 reaction energy and barrier height for the (7,4) slab model using \cref{eq:tio_mp2,eq:tio_hf_compcorr,eq:tio_mp2c_compcorr}.}
        \label{tab:tio_mp2_compcorr}
        % \begin{tabular}{ccc}
        %     \hline\hline
        %     E^{(7,4)}_{\mathrm{HF}}(\mathrm{DZ}/1\times 1\times 1) &    \\
        %     \Delta E^{(6,4)^*}_{\mathrm{HF}}(\mathrm{TZ}/1\times 1\times 1) &   \\
        %     \Delta E^{(6,4)^*}_{\mathrm{HF}}(\mathrm{DZ}/2\times 2\times 1) &   \\
        %     \hline
        %     E^{(7,4)}_{\mathrm{MP2,c}}(\mathrm{DZ}/1\times 1\times 1) &     \\
        %     \Delta E^{(6,4)^*}_{\mathrm{MP2,c}}(\mathrm{TZ}/1\times 1\times 1) &    \\
        %     \hline
        % \end{tabular}
        \begin{tabular}{llllccc}
            \hline\hline
            & Slab & Basis set & BZ sampling &
                $E_{\textrm{ads}}$ & $\Delta E$ & $\Delta E^{\ddagger}$   \\
            \hline\hline
            \multirow{7}*{HF}
            & (7,4) & DZ & $1\times1\times1$ & -15.8 & -6.1 & 9.1 \\
            & $(6,4)^*$ & DZ & $1\times1\times1$ & -15.6 & -5.6 & 9.2 \\
            & $(6,4)^*$ & TZ & $1\times1\times1$ & -18.6 & -7.1 & 8.4 \\
            & $(6,4)^*$ & DZ & $2\times2\times1$ & -7.7 & -4.6 & 9.6 \\
            \cmidrule(lr){2-7}
            & $(6,4)^*$ & $\Delta$TZ & $1\times1\times1$ & 1.2 & -1.4 & -0.8 \\
            & $(6,4)^*$ & DZ & $\Delta2\times2\times1$ & 6.7 & 1.0 & 0.3 \\
            \cmidrule(lr){2-7}
            & (7,4) & TZ & $\Delta2\times2\times1$ & \CG{}-10.9 & \CG{}-6.5 & \CG{}8.6 \\
            \hline\hline
            \multirow{5}*{MP2,c}
            & (7,4) & DZ & $1\times1\times1$ & -17.6 & 6.0 & 1.9   \\
            & $(6,4)^*$ & DZ & $1\times1\times1$ & -17.4 & 6.2 & 2.0   \\
            & $(6,4)^*$ & TZ & $1\times1\times1$ & -17.7 & 7.7 & 0.8   \\
            \cmidrule(lr){2-7}
            & $(6,4)^*$ & $\Delta$TZ & $1\times1\times1$ & -0.4 & 1.4 & -1.3   \\
            \cmidrule(lr){2-7}
            & (7,4) & TZ & $1\times1\times1$ & \CG{}-17.2 & \CG{}7.4 & \CG{}0.7   \\
            \hline\hline
            MP2 & (7,4) & TZ & $2\times2\times1$ & \CG{}-25.6 & \CG{}0.9 & \CG{}9.3   \\
            \hline
        \end{tabular}
    \end{table}

\begin{figure}[!h]
        \centering
        \includegraphics[width=1.0\linewidth]{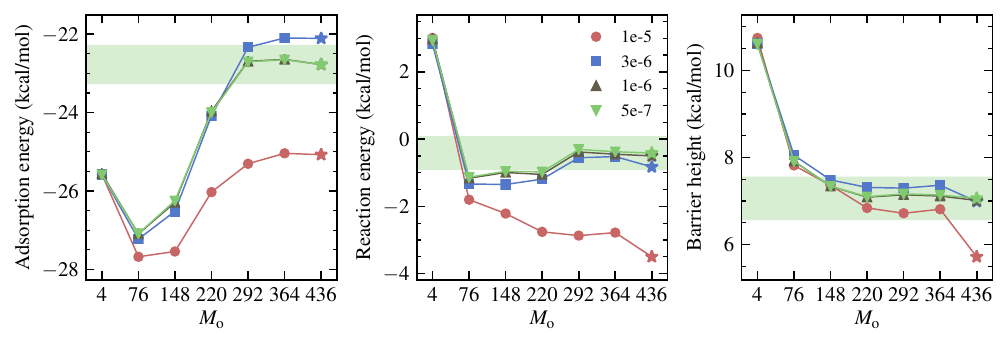}
        \captionsetup{width=\linewidth}
        \caption{\revision{Convergence of the LNO-CCSD(T) adsorption energy (left), reaction energy (middle), and barrier height (right) for water-\tio{} with respect to the embedded cluster size $M_{\mathrm{o}}$.
        The $(6,4)^*$/$1 \times 3$ surface model and the TZ basis set are employed, with a MP2 composite correction evaluated using the $(7,4)$/$1 \times 3$ surface model.
        Data of different color correspond to different virtual LNO truncation parameter $\lambda_{\mathrm{v}}$ (the occupied truncation parameter is chosen to be $\lambda_{\mathrm{o}} = 10\lambda_{\mathrm{v}}$ in all cases except for $\lambda_{\mathrm{v}} = 5 \times 10^{-7}$, where $\lambda_{\mathrm{o}} = 10^{-5}$).
        The range of $\pm 0.5$~kcal/mol from the most converged number (i.e.,~$\lambda_{\textrm{v}} = 5 \times 10^{-7}$ and $M_{\textrm{o}} = 436$ is highlighted by the green shaded area.)
        }
        }
        % \caption{Convergence of the LNO-CCSD(T) reaction energy (left) and barrier (right) for water-\tio{} with respect to the embedded cluster size $M_{\mathrm{o}}$.
        % Data of different color correspond to different virtual LNO truncation parameter $\lambda_{\mathrm{v}}$ (the occupied truncation parameter is chosen to be $\lambda_{\mathrm{o}} = 10\lambda_{\mathrm{v}}$ in all cases except for $\lambda_{\mathrm{v}} = 5\mathrm{e}-7$, where $\lambda_{\mathrm{o}} = 1\mathrm{e}-5$).}
        \label{fig:tio_lnoconv}
    \end{figure}

    \noindent
    \textbf{LNO-CCSD(T)}.
    We calculate the LNO-CCSD(T) correlation energy for the $(6,4)^*$ slab model using \cref{eq:Ec_lnoemb} with a MP2 composite correction using the MP2 energy for the $(7,4)$ slab model calculated by \cref{eq:tio_mp2}.
    The convergence of the LNO-CCSD(T) reaction energetics with respect to the cluster size $M_{\mathrm{o}}$ is shown in \cref{fig:tio_lnoconv}, where the choice of $M_{\mathrm{o}}$ follows that in water-\alo{}, i.e.,~the first $4$ LOs are localized on water, followed by $72$ LOs in each atomic layer.
    \Cref{fig:tio_lnoconv} suggests that for all choices of the LNO truncation parameters, $M_{\mathrm{o}} = 292$, i.e.,~including LOs in up to the fourth atomic layer is sufficient for a converged results.
    The converged energetics from using different LNO truncation parameters (marked as stars in \cref{fig:tio_lnoconv}) are then plotted in Fig.~\fakeref{M3C} to investigate the convergence with respect to the LNO truncation, where we also see a fast convergence to the desired accuracy of $1$~kcal/mol, as already discussed in the main text.

    \revision{
    \subsection{Infinite vacuum correction}
    \label{subsec:tio_dipcorr}

    \noindent
    \textbf{PBE+D3, PBE+$U$+D3, and SCAN+D3}.
    We use the dipole correction at PBE level as implemented in Quantum Espresso to account for the finite-vacuum effect.
    For the $(7,4)$ slab model, we obtained $\Delta_{\infty} = 0.20$, $-0.03$ and $-0.09$~kcal/mol for the adsorption energy, reaction energy, and barrier height, respectively, as also tabulated in \cref{tab:tio_final}.
    These corrections are also applied to obtain the final PBE+$U$+D3 and SCAN+D3 energetics in \cref{tab:tio_final}.

    \noindent
    \textbf{MP2 and LNO-CCSD(T)}.
    For MP2, we manually increase $z_{\textrm{max}}$ and repeat the calculations using the $(6,4)^*$ model and the DZ basis set.
    The resulting energies are plotted in \cref{fig:tio_mp2_vacconv}, where extrapolation to the infinite vacuum limit using \cref{eq:alo_vacextrap} is also shown.
    From the plot, we obtain $\Delta_{\infty} = 0.51$, $0.07$, and $-0.15$~kcal/mol for $E_{\textrm{ads}}$, $\Delta E$, and $E^{\ddagger}$, respectively, which are also listed in \cref{tab:tio_final}.
    In the spirit of the MP2 composite correction, these corrections are also applied to obtain the final LNO-CCSD(T) energetics in \cref{tab:tio_final}.

    \begin{figure}[!h]
        \centering
        \includegraphics[width=1.0\linewidth]{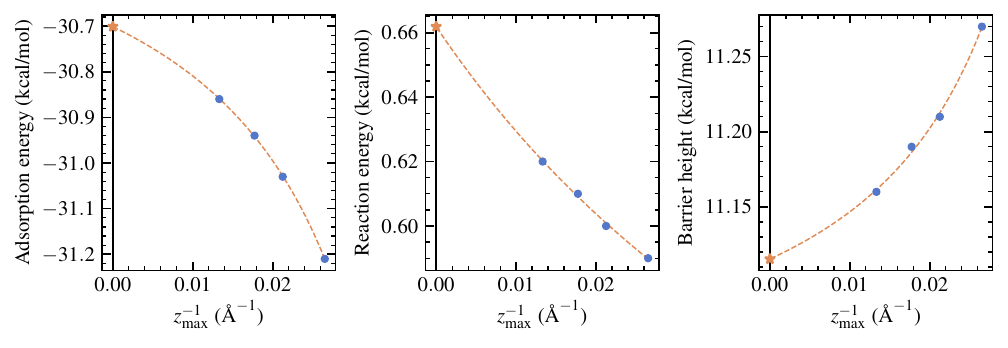}
        \caption{Convergence of the MP2 adsorption energy (left), reaction energy (middle), and barrier height (right) for water-\tio{} as a function of the reciprocal dimension perpendicular to the surface.
        Extrapolations based on \cref{eq:alo_vacextrap} are shown in orange.
        }
        \label{fig:tio_mp2_vacconv}
    \end{figure}
    }

    \subsection{Vibrational corrections}
    \label{subsec:tio_zpe}

    \revision{%
    DFPT calculations at PBE level are performed for the clean surface, molecular adsorption, dissociated adsorption, and transition state of a (4,2)/$1 \times 3$ surface model.
    Atoms in the three layers from the bottom are kept frozen in the DFPT calculations to reduce the computational cost.
    The reciprocal space is sampled at the $\Gamma$-point in the DFPT calculations.
    The ZPE corrections for the adsorption energy, reaction energy, and barrier height are already listed in \cref{tab:tio_final}.
    }

    \clearpage

    \section{Optimized geometries}

    \subsection{Bulk lattice constants}
    \label{subsec:bulk_lat_const}

    \begin{table}[!h]
        \centering
        \caption{Optimized lattice constants for bulk $\alpha$-\ce{Al2O3} and rutile \ce{TiO2}.
        The PBE geometries are used for the subsequent surface calculations.
        PBE+D3 geometries are also included for comparison.
        For both crystals, the difference between PBE and PBE+D3 lattice constants is smaller than 0.4\%.
        $a/b/c$ and $\alpha/\beta/\gamma$ are reported in $\textrm{\AA}$ and degree, respectively.}
        \begin{tabular}{lccccccc}
            \hline\hline
            \multicolumn{7}{c}{$\alpha$-\ce{Al2O3}}  \\
            \hline
            & $a$ & $b$ & $c$ & $\alpha$ & $\beta$ & $\gamma$ \\
            \hline
            PBE &
                4.813 & 4.813 & 13.133 & 90 & 90 & 120  \\
            PBE+D3 &
                4.796 & 4.796 & 13.085 & 90 & 90 & 120  \\
            Deviation &
                $0.35\%$ & $0.35\%$ & $0.37\%$ & & &    \\
            \hline
            \multicolumn{7}{c}{rutile \ce{TiO2}}  \\
            \hline
            & $a$ & $b$ & $c$ & $\alpha$ & $\beta$ & $\gamma$ \\
            \hline
            PBE &
                4.649 & 4.649 & 2.971 & 90 & 90 & 90  \\
            PBE+D3 &
                4.636 & 4.636 & 2.963 & 90 & 90 & 90  \\
            Deviation &
                $0.28\%$ & $0.28\%$ & $0.27\%$ & & &    \\
            \hline
        \end{tabular}
    \end{table}

    \subsection{Optimized geometries}
    \label{subsec:opt_geom}

    The PBE+D3 optimized surface geometries can be found in \url{https://github.com/hongzhouye/supporting_data/tree/main/2023/arXiv%3A2309.14640}.

    \subsection{Energy uncertainty from the geometry}
    \label{subsec:est_geom_uncertainty}

    To estimate the uncertainty of the employed PBE+D3 geometry, we repeated the surface geometry relaxation using revPBE~\cite{Zhang98PRL}+D3 and compare the adsorption and reaction energetics evaluated using both the PBE+D3 optimized geometries and the revPBE+D3 optimized geometries.
    The results are listed in \cref{tab:geom_uncertainty_revpbe}.
    We see that in both cases, the difference in the calculated adsorption and reaction energetics due to using the two sets of geometries is well-below chemical accuracy.

    \begin{table}
        \centering
        \caption{revPBE+D3 adsorption and reaction energetics evaluated using PBE+D3-optimized geometries and revPBE+D3-optimized geometries.
        Results from the finite-vacuum geometries are shown without further ZPE or infinite vacuum corrections.
        All numbers are reported in kcal/mol.}
        \label{tab:geom_uncertainty_revpbe}
        \begin{tabular}{llccc}
            \hline\hline
            & Geometry & $E_{\textrm{ads}}$ & $\Delta E$ & $\Delta E^{\ddagger}$   \\
            \hline
            \multirow{2}*{\alosurf{}}
                & PBE+D3 & $-27.17$ & $-10.23$ & $5.39$    \\
                & revPBE+D3 & $-27.22$ & $-10.22$ & $5.43$    \\
            \hline
            \multirow{2}*{\tiosurf{}}
                & PBE+D3 & $-21.63$ & $0.85$ & $7.92$    \\
                & revPBE+D3 & $-21.93$ & $1.05$ & $8.10$    \\
            \hline
        \end{tabular}
    \end{table}

    \section{Timing}

    \revision{The CPU cost for a typical single-point calculation using different methods for water-\alo{} and water-\tio{} is tabulated in \cref{tab:alo_timing} and \cref{tab:tio_timing}.
    }

    \begin{table}[!h]
        \centering
        \caption{CPU cost (unit: hour) for a typical single-point calculation of the water-\alo{} system measured using 16 CPU cores on a single node.}
        \label{tab:alo_timing}
        \begin{tabular}{lllllr}
            \hline\hline
            Method & Code & Basis & Slab model & $k$-point & CPU cost \\
            \hline
            PBE+D3 & QE & 100 Ry & $(12,8)$/$2\times2$ & $1\times1\times1$ & $0.3$ \\
            SCAN+D3 & QE & 150 Ry & $(12,8)$/$2\times2$ & $1\times1\times1$ & $0.5$ \\
            PBE0+D3 & PySCF & TZ & $(12,8)$/$2\times2$ & $1\times1\times1$ & $42.2$ \\
            HF & PySCF & TZ & $(12,8)$/$2\times2$ & $1\times1\times1$ & $30.6$ \\
            MP2 & PySCF & TZ & $(12,8)$/$2\times2$ & $1\times1\times1$ & $7.4$ \\
            LNO-CCSD(T) & PySCF & TZ & $(12,8)$/$2\times2$ & $1\times1\times1$ & $403.4$ \\
            \hline
        \end{tabular}
    \end{table}

    \begin{table}[!h]
        \centering
        \caption{CPU cost (unit: hour) for a typical single-point calculation of the water-\tio{} system measured using 16 CPU cores on a single node.
        For HF and MP2, CPU time for the most expensive calculation in the respective protocols is listed.}
        \label{tab:tio_timing}
        \begin{tabular}{lllllr}
            \hline\hline
            Method & Code & Basis & Slab model & $k$-point & CPU cost \\
            \hline
            PBE+D3 & QE & 100 Ry & $(7,4)$/$1\times3$ & $2\times2\times1$ & $4.7$ \\
            PBE+$U$+D3 & QE & 100 Ry & $(7,4)$/$1\times3$ & $2\times2\times1$ & $4.3$ \\
            SCAN+D3 & QE & 150 Ry & $(7,4)$/$1\times3$ & $2\times2\times1$ & $49.0$ \\
            HF & PySCF & DZ & $(6,4)^*$/$1\times3$ & $2\times2\times1$ & $1141.9$ \\
            MP2 & PySCF & TZ & $(6,4)^*$/$1\times3$ & $1\times1\times1$ & $113.7$ \\
            LNO-CCSD(T) & PySCF & TZ & $(6,4)^*$/$1\times3$ & $1\times1\times1$ & $19099.9$ \\
            \hline
        \end{tabular}
    \end{table}

    \clearpage

    \bibliography{refs_si}